\documentclass{JHEP3}
\usepackage{amsmath}
\usepackage{epsfig,multicol,bbm}
\usepackage{rotating}
\providecommand{\U}[1]{\protect\rule{.1in}{.1in}}
\newcommand{\mytitle}{\title}
\newcommand{\myauthor}{\author}
\newcommand{\be} {\begin{equation}}
\newcommand{\ee} {\end{equation}}
\newcommand{\bea}{\begin{eqnarray}}
\newcommand{\eea}{\end{eqnarray}}

\newcommand{\xp}{\ensuremath{x_{I\!\!P}}}

\def\hbar{{\bar h}}

\mytitle{Further success of the colour dipole model}
\myauthor{J.R. Forshaw${}^1$, R. Sandapen${}^2$ and G. Shaw${}^1$ \\
${}^1$School of Physics \& Astronomy, University of Manchester, \\
Oxford Road, Manchester M13 9PL, U.K.\\
${}^2$D\'epartement de Physique et d'Astronomie, Universit\'e de Moncton, Moncton,
N-B. E1A 3E9. Canada.\\ \email{forshaw@mail.cern.ch}}
\preprint{MAN/HEP/2006/23}
\abstract{We confront a very wide body of HERA diffractive electroproduction data with
the predictions of the colour dipole model. We focus upon three different parameterisations
of the dipole scattering cross-section and find good agreement for all observables. There
can be no doubting the success of the dipole scattering approach and more precise observations
are needed in order to expose its limitations.}

\begin{document}

\section{Introduction}

\begin{figure}[tbh]
\begin{center}
\includegraphics[width=8cm]{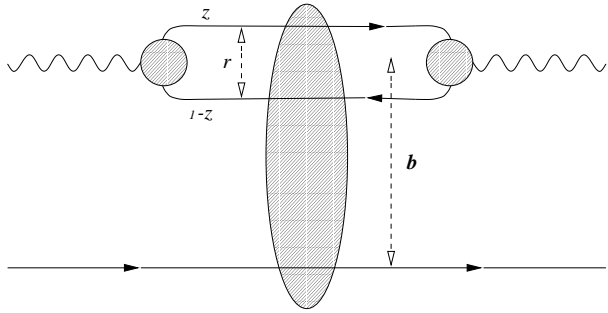}
\end{center}
\caption{The colour dipole model for elastic Compton scattering $\gamma^{\ast
}p\rightarrow\gamma^{\ast}p$.}%
\label{f2dipole}%
\end{figure}

In the colour dipole model ~\cite{NZ91,Mueller94}, the forward amplitude for virtual Compton
scattering is assumed to be dominated by the mechanism illustrated in
Fig.\ref{f2dipole} in which the photon fluctuates into a $q\bar{q}$ pair of
fixed transverse separation $r$ and the quark carries a fraction $z$ of the
incoming photon light-cone energy. Using the Optical Theorem, this leads to
\begin{equation}
\sigma_{\gamma^{\ast}p}^{L,T}=\int dz\;d^{2}r\ |\Psi_{\gamma}^{L,T}%
(r,z,Q^{2})|^{2}\sigma(s^{\ast},r)\label{dipoledis}
\end{equation}
for the total virtual photon-proton cross-section, where $\Psi_{\gamma}^{L,T}$
are the appropriate spin-averaged light-cone wavefunctions of the photon and
$\sigma(s^{\ast},r)$ is the dipole cross-section. The dipole cross-section is
usually assumed to be independent of $z$, and is parameterised in terms of an
energy variable $s^{\ast}$ which depends upon the model.

Thus using Eq.(\ref{dipoledis}) we are able to compute the deep inelastic structure
function $F_2(x,Q^2)$. The power of the dipole model formulation lies in the fact that
the same dipole cross-section $\sigma(s^*,r)$ appears in a variety of other 
observables which involve the scattering of a real or virtual photon off a 
hadronic (or nuclear) target at high centre-of-mass (CM) energy. The largeness of the
CM energy guarantees the factorization of scattering amplitudes into a product of
wavefunctions and a universal dipole cross-section. In this paper we wish to
test the universality of the dipole cross-section using a wide range of high
quality data collected at the HERA $ep$ collider. Moreover, we also wish to examine
the extent to which to data are able to inform us of the role, if any, played by
non-linear saturation dynamics.

Specifically, we shall consider three particular parameterisations of the dipole
cross-section which have been presented in the literature and which will
be discussed in more detail below. All three have
been constrained by fitting only to the HERA data on $F_2(x,Q^2)$ and so they can be
used to make genuine predictions for other observables. In this paper we will
compare those predictions to data on the charm structure function $F_2^c(x,Q^2)$,
the cross-section for Deeply Virtual Compton Scattering (DVCS), the cross-section
for diffractive $J/\Psi$ production and the diffractive structure function $F_2^{D(3)}$.

In the first instance our results confirm the validity of the dipole model: it does
make sense to speak of a universal dipole cross-section which is able to
account for both soft and hard diffraction in a wide variety of photo-processes. 
We also attempt to ascertain the extent to which the new data are able to discriminate
between the predictions of the three dipole models we use. These models
can perhaps best be described as parameterisations which incorporate certain general 
theoretical ideas and can be viewed as providing an indication 
of the uncertainties remaining in the dipole cross-section once the precise structure 
function data have been accounted for. Unfortunately we shall find that the new data are 
not quite precise enough to discriminate between the models or to add any significant
evidence for saturation beyond that already present in the $F_2$ data. We do however
find that in diffractive photo/electro-production progress is hindered by the
lack of a precise enough measurement of the forward slope parameter $B$ which 
determines the $t$-dependence of the final state proton. We take this parameter from
data and its error translates into an uncertainty on the normalisation of the predicted
cross-sections. A more precise measurement of this quantity would provide a significant
additional constraint.

It is clear that having extracted the dipole cross-section from data, one would hope
eventually to explain the result using QCD. Unfortunately, while QCD in its present
state of development is able to suggest qualitative features of the dipole cross-section,
more quantitative predictions are not possible without severe approximations. To remedy
this situation is a challenge which lies beyond the scope of this paper.

\section{The dipole cross-section}

We now turn to the three different models used to describe the dipole cross-section.
Before doing so however, we shall first discuss our choice of photon wavefunction.
For small $r$, the light-cone photon wavefunctions are given by the tree level
QED expressions~\cite{NZ91}
\begin{align}
&  \hspace*{-1.3cm}|\Psi_{\gamma}^{L}(r,z,Q^{2})|^{2}=\frac{6}{\pi^{2}}%
\alpha_{\text{em}}\sum_{f=1}^{n_{f}}e_{f}^{2}Q^{2}z^{2}(1-z)^{2}K_{0}%
^{2}(\epsilon r)\label{eq:psi^2}\\
&  \hspace*{-1.3cm}|\Psi_{\gamma}^{T}(r,z,Q^{2})|^{2}=\frac{3}{2\pi^{2}}%
\alpha_{\text{em}}\sum_{f=1}^{n_{f}}e_{f}^{2}\left\{  [z^{2}+(1-z)^{2}%
]\epsilon^{2}K_{1}^{2}(\epsilon r)+m_{f}^{2}K_{0}^{2}(\epsilon r)\right\}
\end{align}
where
\begin{equation}
\epsilon^{2}=z(1-z)Q^{2}+m_{f}^{2}\ \;.\label{epsilon}
\end{equation}
Here $K_{0}(x)$ and $K_{1}(x)=-\partial_{x}K_{0}(x)$ are modified Bessel
functions and the sum is over all $n_{f}=4$ quark flavours $f$.
These wavefunctions decay exponentially at large $r$, with typical $r$-values
of order $Q^{-1}$ at large $Q^{2}$ and of order $m_{f}^{-1}$ at $Q^{2}=0$.
However for large dipoles $r\gtrsim1$ fm, which are important at low $Q^{2}$,
a perturbative treatment is not really appropriate. In this region some
authors~\cite{FKS99} modify the perturbative wavefunction by an enhancement
factor motivated by generalised vector dominance (GVD)
ideas~\cite{GVD1,GVD2,FGS98}, while others~\cite{GW99a} achieve a similar but
broader enhancement by varying the quark mass\footnote{For a fuller discussion
of these points see \cite{FS06}.}. In practice \cite{FS04}, the difference
between these two approaches only becomes important when analysing the precise
real photoabsorption data from fixed-target experiments \cite{Caldwell78}.
Since we will not consider these data here, we will adopt the simpler practise
of using a perturbative wavefunction at all $r$-values, and adjusting the
quark mass to fit the data.

Turning now to the dipole cross-section, all three models are 
consistent with the physics of colour transparency for small dipoles and exhibit
soft hadronic behaviour for large dipoles. As stated above, the model parameters 
are determined by fitting only to the DIS structure function data. The resulting 
dipole cross-sections can then be used to make genuine predictions for other reactions. 
Since the details of all three models have been published elsewhere, we shall 
here summarise their properties only rather briefly. 

\subsection{The FS04 Regge model}

This simple model \cite{FS04} combines colour transparency for small dipoles
$r<r_{0}$ with \textquotedblleft soft pomeron\textquotedblright\ behaviour for
large dipoles $r>r_{1}$ by assuming
\begin{align}
\sigma(x_{m},r) &  =A_{H}r^{2}x_{m}^{-\lambda_{H}}~~{\mathrm{for}}
~~r<r_{0}~~{\mathrm{and}}\nonumber\\
&  =A_{S}x_{m}^{-\lambda_{S}}~~{\mathrm{for}}~~r>r_{1},\label{eq:FS04}
\end{align}
where
\begin{equation}
x_{m}=\frac{Q^{2}}{Q^{2}+W^{2}}\left(  1+\frac{4m_f^{2}}{Q^{2}}\right).
\label{eq:xmod}
\end{equation}
For light quark dipoles, the quark mass $m_f$ is a parameter in the fit, whilst
for charm quark dipoles the mass is fixed at 1.4~GeV. In the intermediate
region $r_{0}\leq r\leq r_{1}$, the dipole cross-section is given by
interpolating linearly between the two forms of Eq.(\ref{eq:FS04}).

If the boundary parameters $r_{0}$ and $r_{1}$ are kept constant then this
parameterisation reduces to a sum of two powers, as might be predicted in a
two pomeron approach, and can be thought of as an update of the original FKS
Regge model \cite{FKS99} to accommodate the latest data. It is plainly
unsaturated, in that the dipole cross-section obtained at small $r$-values and
fixed $Q^{2}$ grows rapidly with increasing $s$ (or equivalently with
decreasing $x$) without damping of any kind.

\subsection{The FS04 Saturation model}

Saturation can be introduced into the above model by adopting a method
previously utilised in \cite{MFGS2000}. Instead of taking $r_{0}$ to be
constant, it is fixed to be the value at which the hard component is some
specified fraction of the soft component, i.e.
\begin{equation}
\sigma(x_{m},r_{0})/\sigma(x_{m},r_{1})=f
\end{equation}
and $f$ instead of $r_{0}$ is treated as a parameter in the fit. This
introduces no new parameters compared to the Regge model. However, the scale
$r_{0}$ now moves to lower values as $x$ decreases, and the rapid growth of
the dipole cross-section at a fixed, small value of $r$ begins to be damped as
soon as $r_{0}$ becomes smaller than $r$. In this sense we model saturation,
albeit crudely, with $r_{0}$ the saturation radius.

\subsection{The CGC saturation model}

In addition we shall consider the CGC dipole model originally presented by
Iancu, Itakura and Munier \cite{IIM04}. This model aims to include the main
features of the \textquotedblleft Colour Glass Condensate\textquotedblright%
\ regime, and can be thought of as a more sophisticated version of the
original \textquotedblleft Saturation Model\textquotedblright\ of
Golec-Biernat and W\"{u}sthoff \cite{GW99a}. Since the original Iancu et al
dipole cross-section was obtained after a three flavour fit to the DIS data it
is not well suited to making predictions for processes involving charm quarks.
Consequently, we instead use a new four-flavour CGC fit due to 
Kowalski, Motyka and Watt \cite{KMW}.

\vspace*{1.5cm}

The parameters of the FS04 models were determined by fitting
the recent ZEUS $F_{2}$ data \cite{ZEUSf2data} in the kinematic range
\begin{equation}
0.045\,\mathrm{GeV}^{2}<Q^{2}<45\,\mathrm{GeV}^{2}\hspace{0.5cm}x\leq0.01\;
\end{equation}
whilst the CGC fit of \cite{KMW} is to data with $Q^2 > 0.25$ GeV$^2$ (the
other limits are as for FS04).
The corresponding H1 data \cite{H1f2data} could also be used, but it would
then be necessary to float the relative normalisation of the two data sets. We
do not do this since the ZEUS data alone suffice. The resulting parameter
values are tabulated in the original papers; we do not reproduce them here,
but confine ourselves to some general comments.

Both the FS04 saturation and the CGC models gave excellent fits to the $F_{2}$
data, while the FS04 Regge fit was not satisfactory, suggesting that
saturation may be required to fit the data~\cite{FS04}.\footnote{The FS04 saturation fit has a 
$\chi^2=155$ for the 156 data points we consider. For the same data, the CGC model has 
$\chi^2=160$ and the FS04 Regge fit has $\chi^2=428$.} 
However this evidence
for saturation depends upon using low-$Q^{2}$ data and disappears if the data
are restricted to $Q^{2}>2$ GeV$^{2}$, whereupon excellent fits can be
obtained in all three models. 

We use the CGC fit with
$\sigma_{0}=35.7$ mb presented in Table 5 of \cite{KMW} and note that although 
the fit is to data with $Q^2 > 0.25$ GeV$^2$ the fit is actually very good
all the way down to $Q^2 = 0.045$ GeV$^2$.

At this point we have three well-determined parameterisations of the colour
dipole cross-section. These can be used to yield predictions for other
processes. In the next sections we shall take a look at Deeply Virtual Compton 
Scattering (DVCS), the charm structure function $F_2^c$, exclusive $J/\Psi$
production and the diffractive structure function $F_2^{D(3)}$.
We always choose to show the Regge fit, even though it does not fit the
$F_{2}$ data particularly well, in order to indicate the discriminatory power
of the data. We stress that in all cases, the photon wavefunctions and dipole
cross-sections are precisely those determined from the fits to $F_{2}$ data,
without any adjustment of parameters.

\section{Exclusive processes}

\begin{figure}[h]
\begin{center}
\includegraphics[width=8cm]{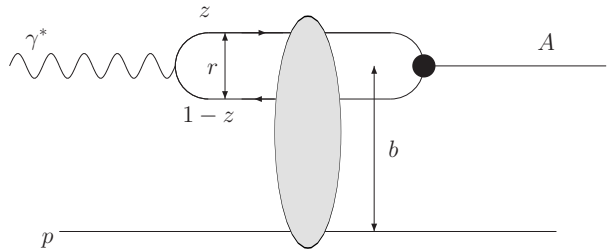}
\end{center}
\caption{The colour dipole model for the exclusive reactions $\gamma^{\ast
}p\rightarrow Ap$.}%
\label{exclusive}%
\end{figure}

We are interested in the exclusive processes
\begin{equation}
\gamma^{\ast}+p\rightarrow A+p\hspace{1.5cm}A=\gamma,\rho,J/\Psi
,\ldots.\label{generic1}%
\end{equation}
In the dipole model they occur via the mechanism of Fig.\ref{exclusive}
and are described by amplitudes which satisfy
\begin{equation}
\mathrm{Im}A_{\lambda}(s,t=0)=s\,\sum_{h,\bar{h}}\int\mathrm{d}
z\,\mathrm{d^{2}}r\;\Psi_{h,\bar{h}}^{A}({r},z)^{\ast}\,\Psi_{h,\bar{h}
}^{\gamma,\lambda}({r},z,Q^{2})\;\sigma(s^*,r)\,\label{result1}
\end{equation}
where $\lambda=L,T$ 
for longitudinal and transverse photons respectively, $h$ and
$\bar{h}$ are the quark and antiquark helicities and $\Psi_{h,\bar{h}}^{A}
({r},z)$ is the light-cone wavefunction of the particle $A$. The forward
differential cross-section is then given by
\begin{equation}
\left.  \frac{d\sigma}{dt}\right\vert _{t=0}=\frac{1}{16\pi s^{2}}|A_{\lambda
}(s,t=0)|^{2}(1+\beta^{2})~,\label{fdcs}
\end{equation}
where the correction from the real part of the amplitude
\[
\beta=\frac{\mathrm{{Re}A_{\lambda}}(s,t=0)}{\mathrm{{Im}A_{\lambda}}(s,t=0)}
\]
can be estimated using dispersion techniques. Predictions for the measured
total cross-sections are then obtained by assuming an exponential
$t$-distribution and integrating over $t$ to obtain
\begin{equation}
\sigma_{L,T}(\gamma^{\ast}p\rightarrow Ap)=\frac{1}{B}\left.  \frac
{d\sigma^{T,L}}{dt}\right\vert _{t=0}\;,\label{sigmatot}
\end{equation}
where the value of the slope parameter $B$ is taken from experiment. 
We refer to \cite{KMW,KT} for models which attempt a more sophisticated treatment of the
dependence upon the momentum transfer, $t$.

The exclusive processes we consider in this paper are DVCS and $J/\Psi$ production.
Light meson production was studied in our previous paper \cite{fss:04a} where we 
found that the dipole model predictions generally agree well with the data modulo 
a rather strong dependence upon the meson wavefunction. For an alternative 
investigation of the link which exists between low $x$ DIS and exclusive processes at 
high energies we refer to \cite{KS}. 

\subsection{Deeply virtual Compton scattering}

In deeply virtual Compton scattering
\begin{equation}
\gamma^{\ast}+p\rightarrow\gamma+p\;,\label{DVCS}
\end{equation}
the final state particle is a real photon and dipole models provide
predictions for the imaginary part of the forward amplitude with no adjustable
parameters beyond those used to describe DIS. To calculate the forward
cross-section a correction for the contribution of the real part of the 
amplitude has to be included. This correction was estimated in 
\cite{MSS02} and found to be less than $\approx 10\%$ and of a similar size 
in different dipole models. Here we
shall estimate the correction using the FS04 Regge model, where the real part
is given by the Regge signature factors. 
Predictions for the measured total cross-sections are then obtained 
using Eq.(\ref{sigmatot}) where the value of the slope parameter $B$ 
is taken from experiment.

\begin{figure}
\begin{center}
\includegraphics*[width=8cm]{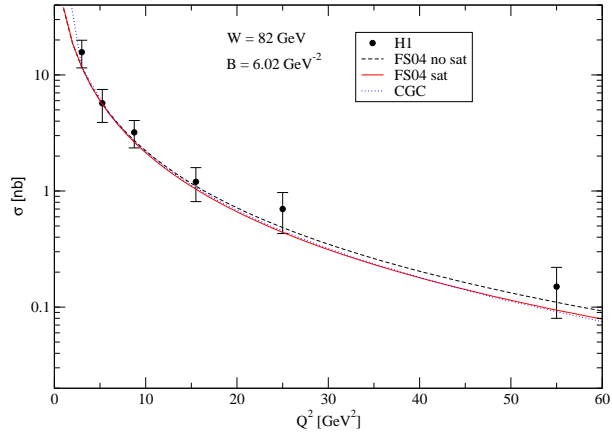}
\end{center}
\caption{ Comparison of the H1 DVCS data \cite{H105} with the predictions of the three
models discussed in the text: $Q^{2}$ dependence at $W=82$ GeV.}%
\label{fig:DVCSH1Q2}%
\end{figure}\begin{figure}
\begin{center}
\includegraphics*[width=8cm]{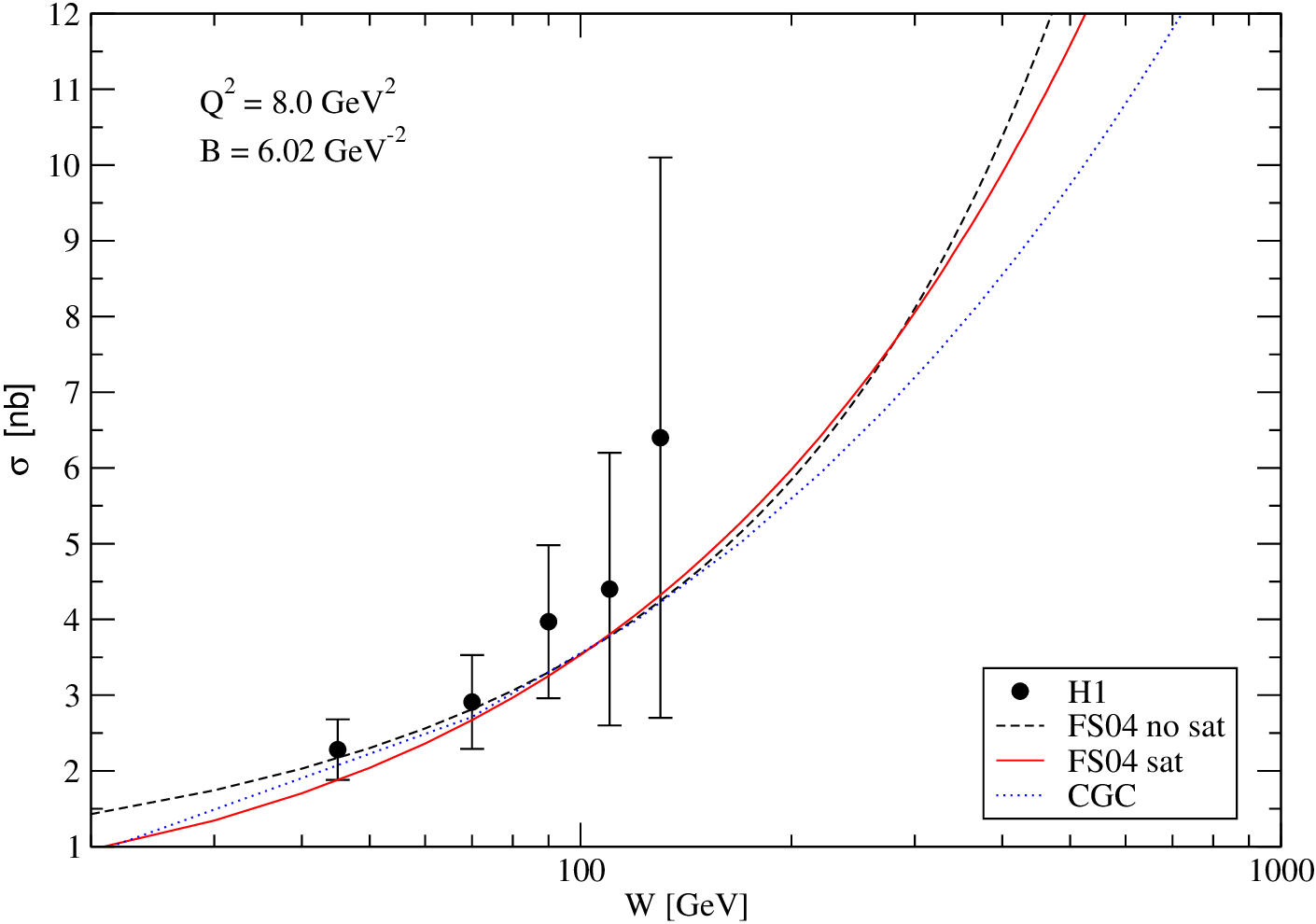}
\end{center}
\caption{ Comparison of the H1 DVCS data \cite{H105} with the predictions of the three
models discussed in the text: $W$ dependence at $Q^2 = 8.0$ GeV$^2$.}%
\label{fig:DVCSH1W}%
\end{figure}

\begin{figure}
\begin{center}
\includegraphics*[width=8cm]{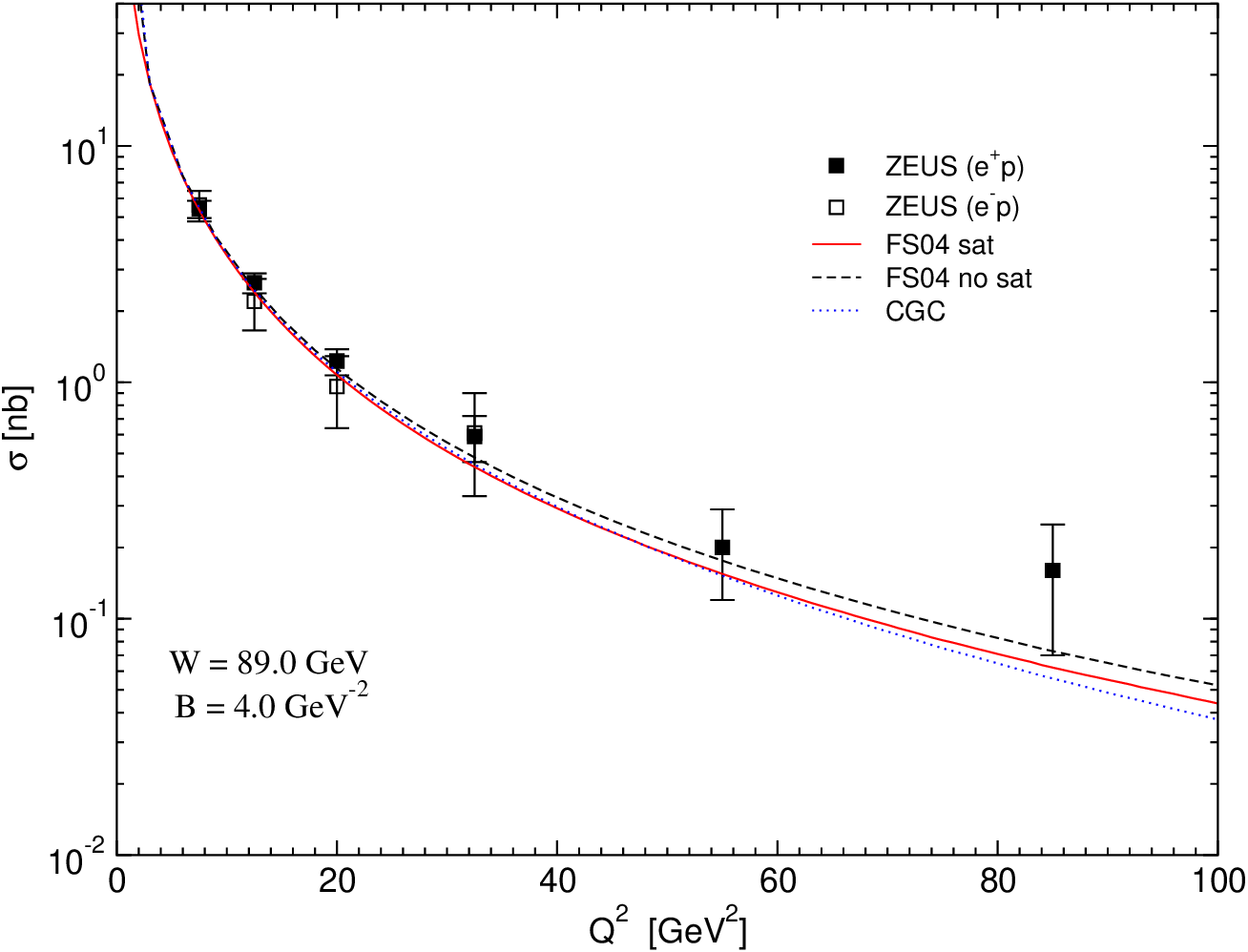}
\end{center}
\caption{ Comparison of the ZEUS DVCS data \cite{zeusdvcs} with the predictions of the three
models discussed in the text: $Q^{2}$ dependence at $W=89$ GeV.}
\label{fig:DVCSZEUSQ2}%
\end{figure}\begin{figure}
\begin{center}
\includegraphics*[width=8cm]{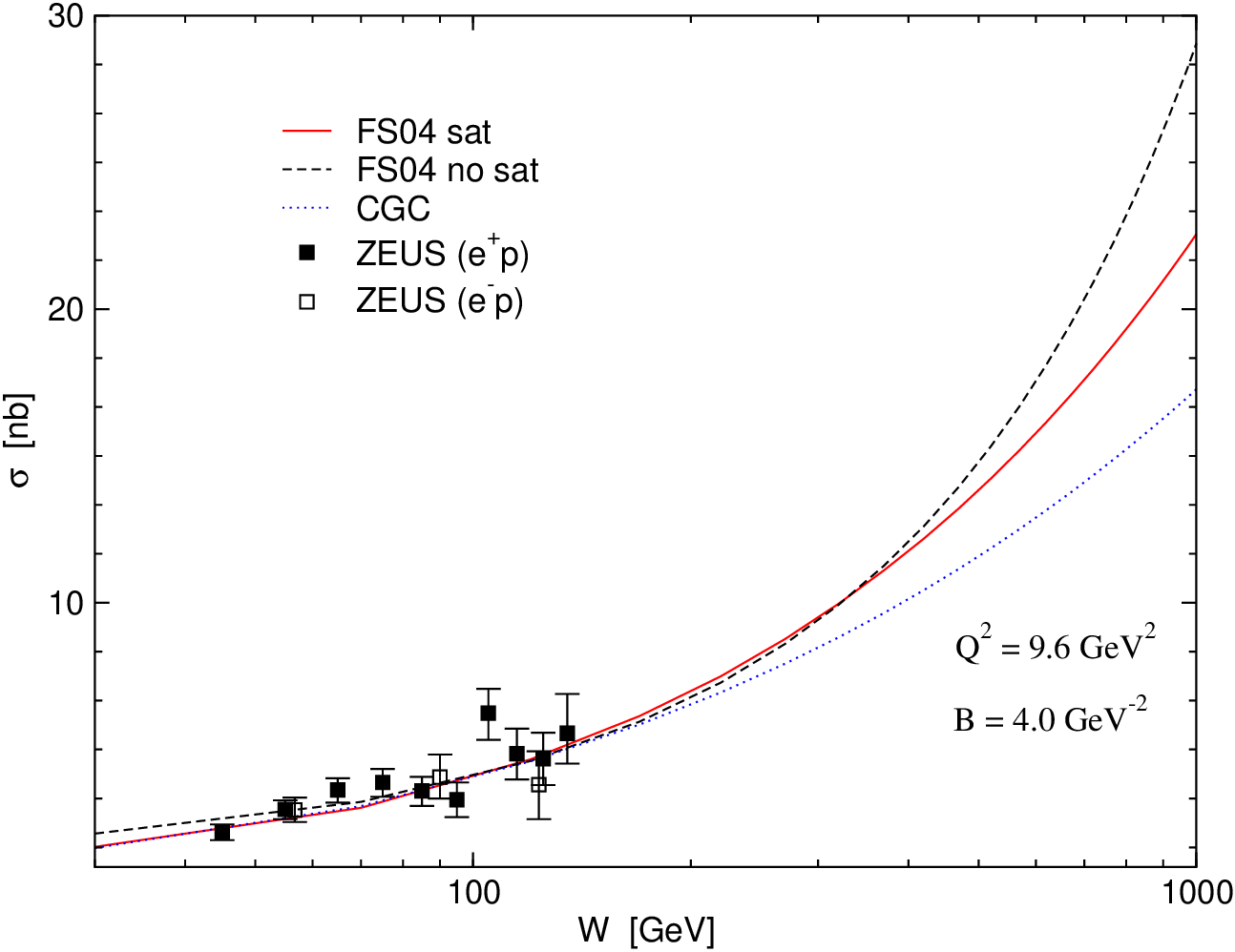}
\end{center}
\caption{ Comparison of the ZEUS DVCS data \cite{zeusdvcs} with the predictions of the three
models discussed in the text: $W$ dependence at $Q^2 = 9.6$ GeV$^2$.}
\label{fig:DVCSZEUSW}%
\end{figure}

The predictions of all three models are compared with the H1 data \cite{H105} 
in Fig.\ref{fig:DVCSH1Q2} and Fig.\ref{fig:DVCSH1W} and with the ZEUS data 
\cite{zeusdvcs} in Fig.\ref{fig:DVCSZEUSQ2} and 
Fig.\ref{fig:DVCSZEUSW}.\footnote{Note that throughout this paper the curves 
labelled `FS04 no sat' correspond to the predictions of the FS04 Regge model.}
For the H1 data we use a fixed value $B=6.02$~GeV$^{-2}$ for the slope
parameter which is in accord with the H1 measurement of $B=6.02\pm0.35\pm
0.39$. For the ZEUS data we take $B=4$~GeV$^{-2}$ which is compatible with
their data. Bearing in mind this normalisation uncertainty, the agreement is good 
for all three models, although significant differences between the
models appear when the predictions are extrapolated to high enough energies,
as one would expect.

\section{$J/\Psi$ and the charm structure function}

We now move on to the predictions for the charm structure function and for
exclusive $J/\Psi$ production
\begin{equation}
\gamma^{\ast}+p \rightarrow J/\Psi + p ~~ .
\label{vmp}
\end{equation}
All three models assume a charm mass of $m_{C}=1.4$ GeV when fitting the
$F_{2}$ data. Since the exclusive process is rather more sensitive to the
charm mass we will allow $m_{c}$ to vary a little without adjusting the
dipole cross-section. This is permissible for small enough variations.

\begin{figure}
\begin{center}
\includegraphics*[width=14cm]{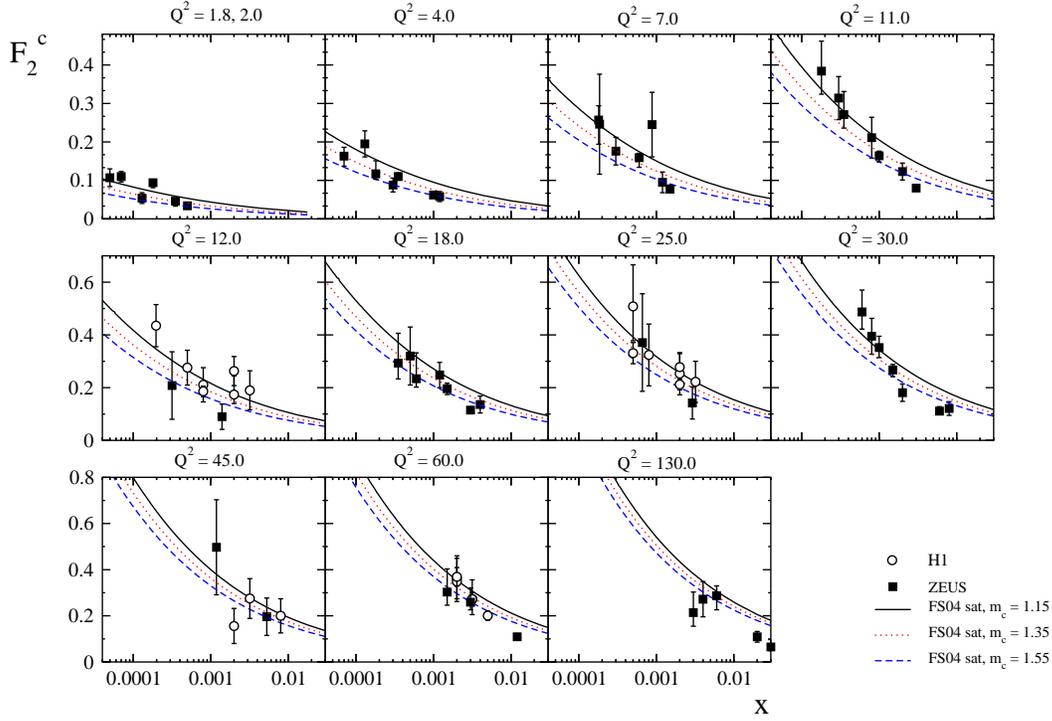}
\end{center}
\caption{ Comparison of the FS04 saturation model predictions for the charmed
structure function $F_{2}^{c}$ with data \cite{F2c_ZEUS,F2c_H1}.}
\label{fig:F2cSAT}
\end{figure}

\begin{figure}
\begin{center}
\includegraphics*[width=14cm]{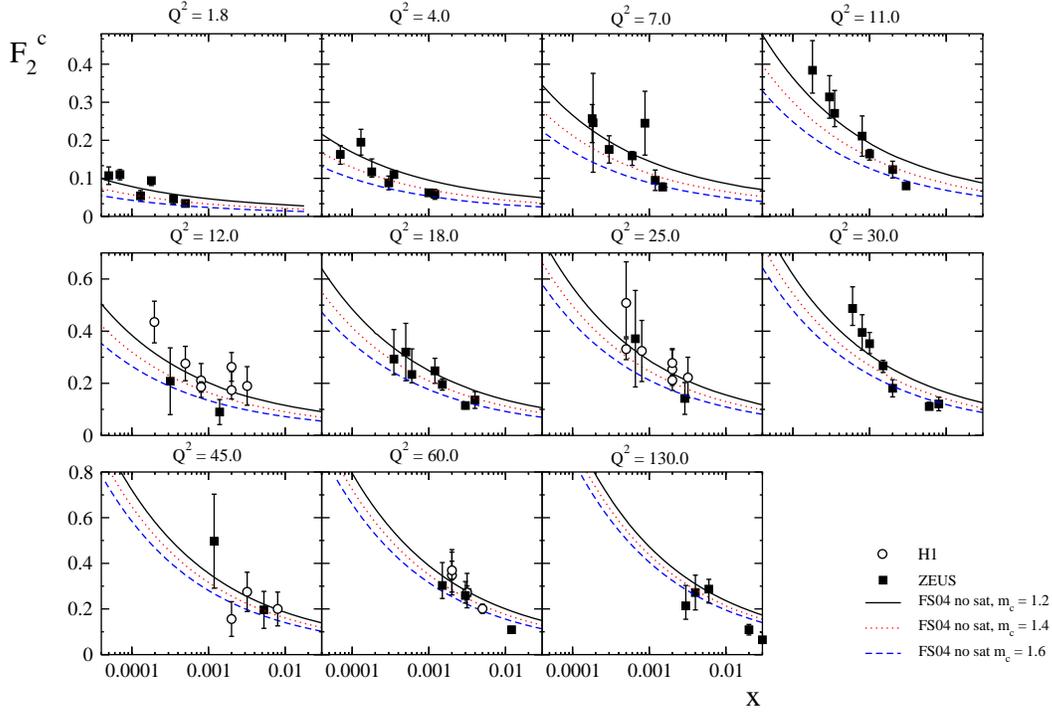}
\end{center}
\caption{ Comparison of the FS04 Regge predictions for the charmed structure
function $F_{2}^{c}$ with data \cite{F2c_ZEUS,F2c_H1}.}
\label{fig:F2cREGGE}
\end{figure}

\begin{figure}
\begin{center}
\includegraphics*[width=14cm]{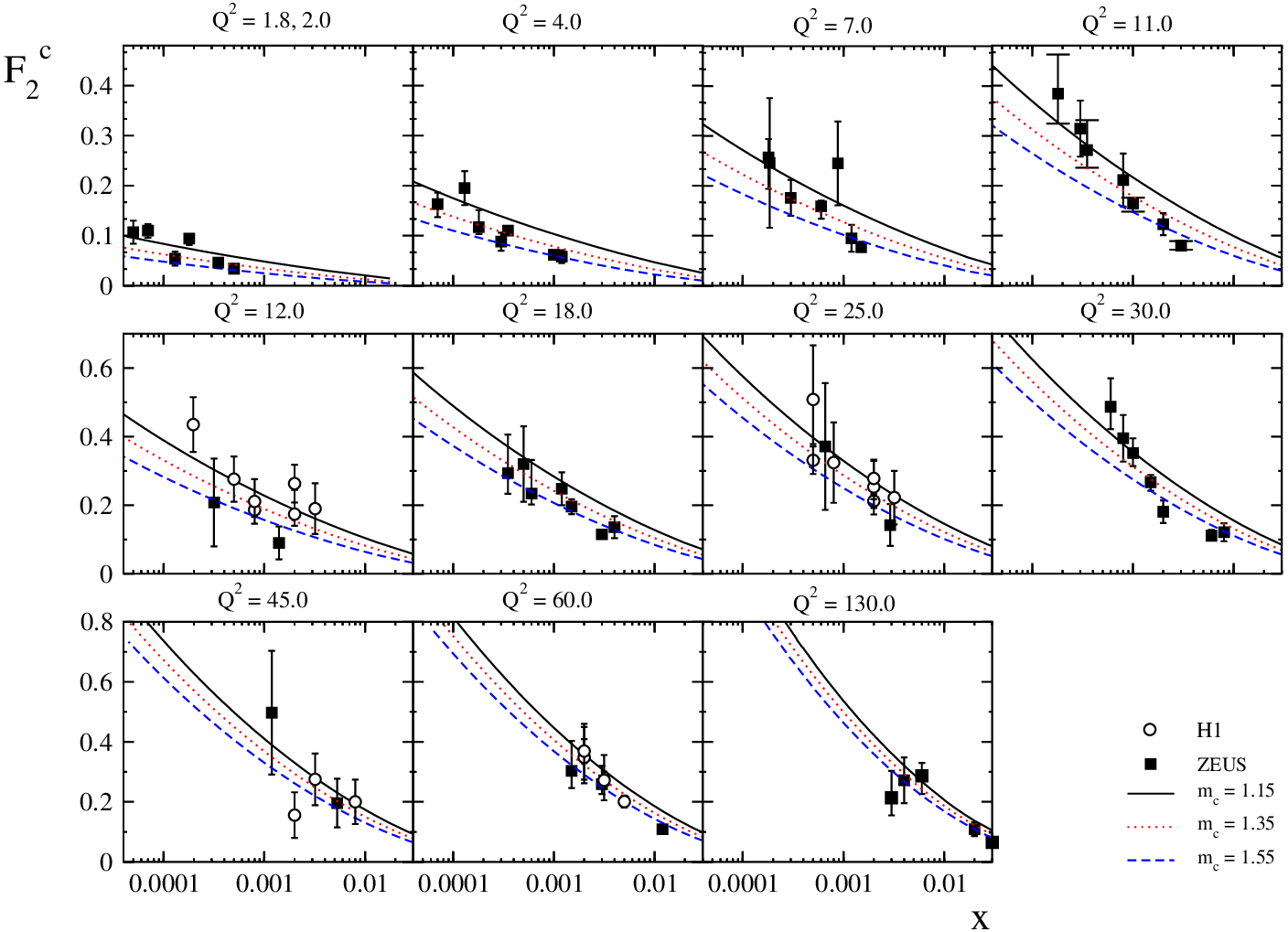}
\end{center}
\caption{ Comparison of the CGC model predictions for the charmed structure
function $F_{2}^{c}$ with data \cite{F2c_ZEUS,F2c_H1}.}
\label{fig:F2cCGC}
\end{figure}

\subsection{The charm structure function}

We begin by discussing the charm structure function $F_{2}^{c}(x,Q^{2})$,
since the results are independent of the vexed question of the vector meson
wavefunction. The charm structure function is given by
\[
F_{2}^{c}(x,Q^{2})=\frac{Q^{2}}{4\pi^{2}\alpha_{\text{em}}}\left(
\vspace*{0.5cm}\sigma_{\gamma^{\ast}p}^{L}+\sigma_{\gamma^{\ast}p}^{T}\right)
\]
where in calculating the total virtual photon-proton cross-sections using
Eq.(\ref{dipoledis}), only the charm component of the light-cone wavefunctions
(\ref{eq:psi^2}) is retained. The resulting predictions are compared to the
ZEUS \cite{F2c_ZEUS} and H1 \cite{F2c_H1} data in Figs.\ref{fig:F2cSAT}--\ref{fig:F2cCGC}. 
A good account of the charm structure function data
can be obtained in all three models by choosing values for the charmed quark mass 
in the reasonable range $1.3\leq m_{c}\leq1.5$ GeV. The key question is
whether one can obtain similarly accurate predictions for the $J/\Psi$
electroproduction data using charm mass values in the same range.

\subsection{ $J/\Psi$ wavefunctions}

\begin{figure}
\begin{center}
\includegraphics*[width=5cm]{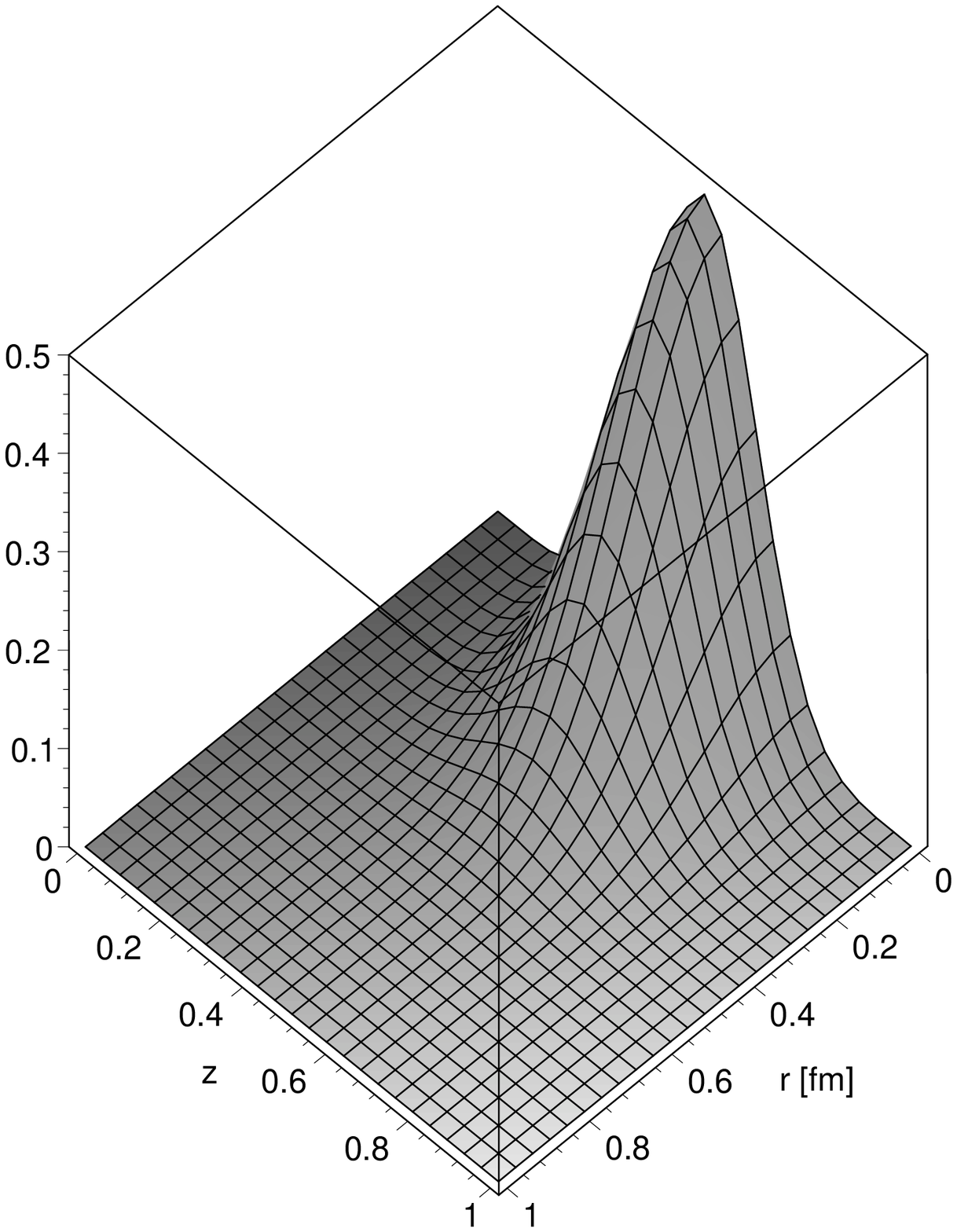} \hspace*{1cm}
\includegraphics*[width=5cm]{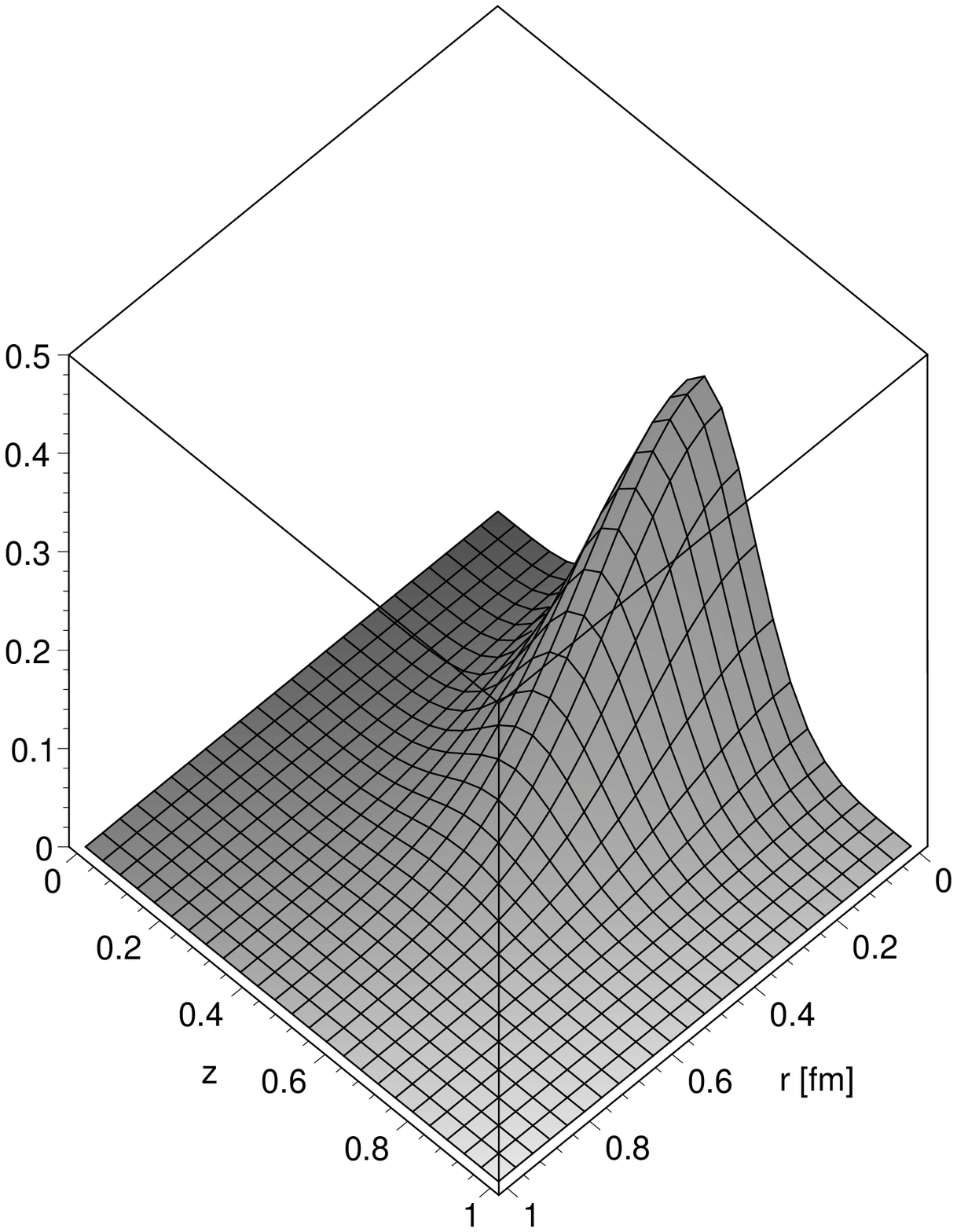}
\end{center}
\caption{The $J/\Psi$-wavefunctions $|\Psi^{L}|^{2}$ (left) and $|\Psi
^{T}|^{2}$ (right) in the DGKP model. }
\label{fig:DGKPJpsi}
\end{figure}

\begin{figure}
\begin{center}
\includegraphics*[width=5cm]{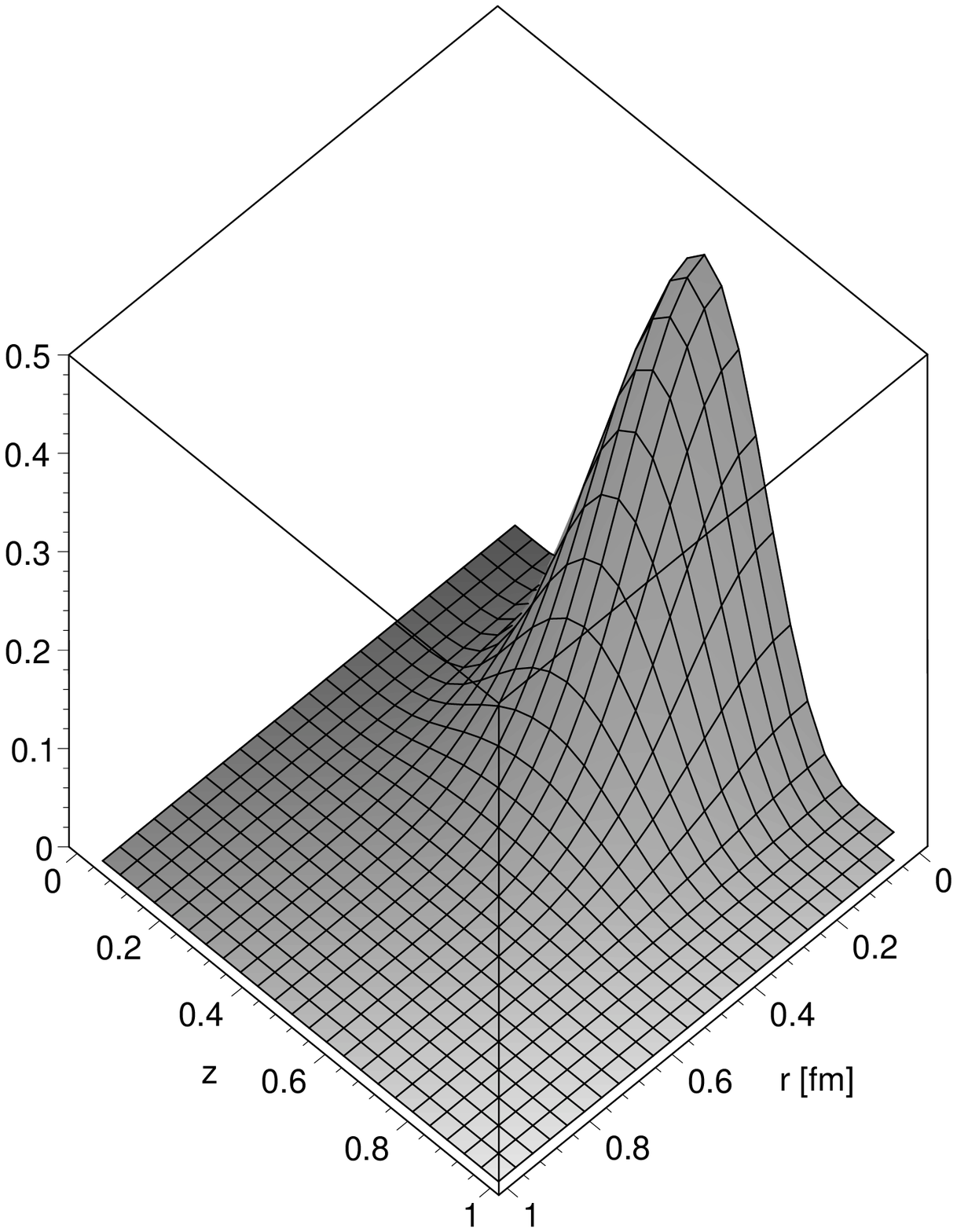} \hspace{1cm}
\includegraphics*[width=5cm]{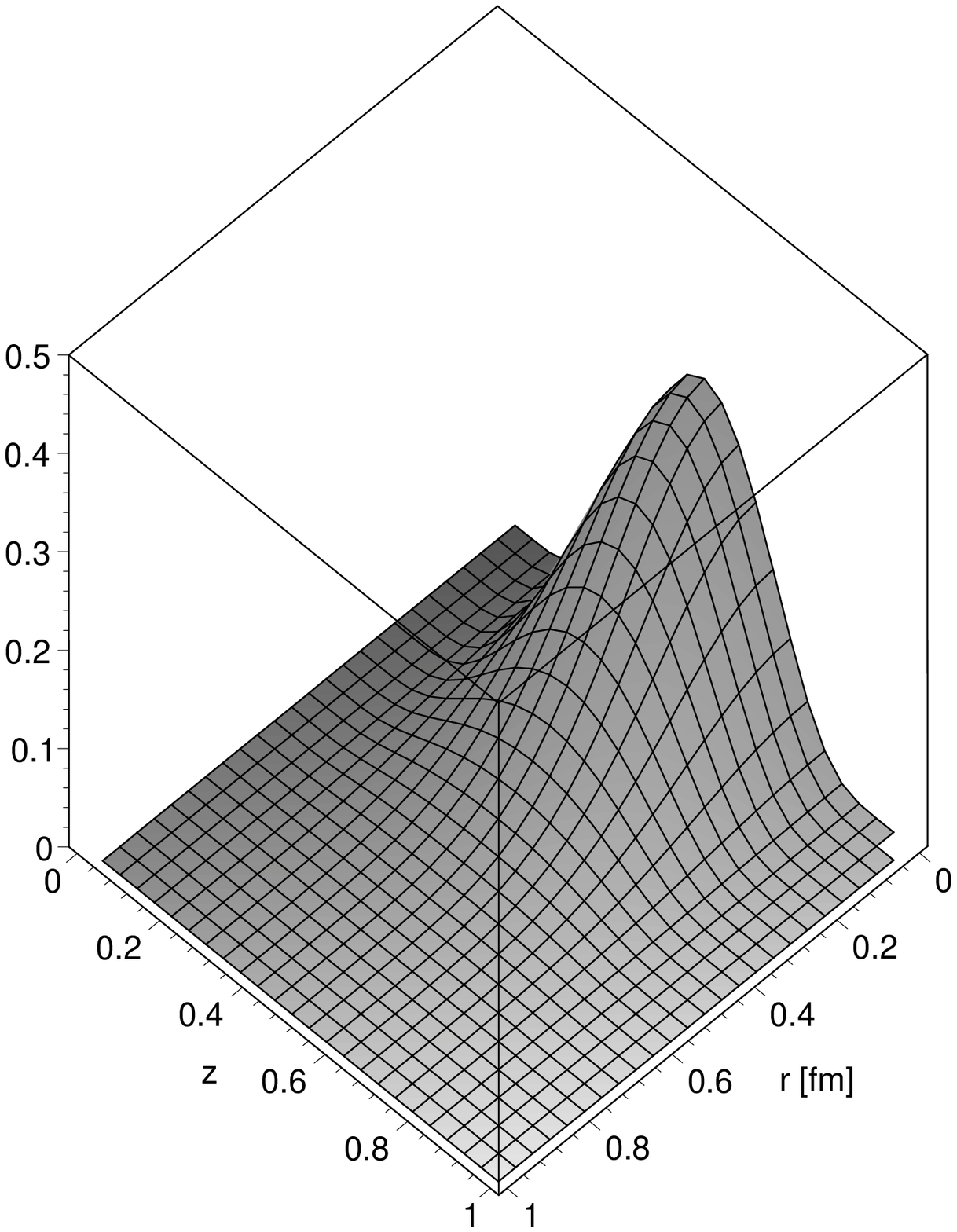}
\end{center}
\caption{The $J/\Psi$-wavefunctions $|\Psi^{L}|^{2}$ (left) and $|\Psi
^{T}|^{2}$ (right) in the Boosted Gaussian model.}
\label{fig:gaussjpsi}
\end{figure}

To calculate vector-meson production we need to know the light-cone
wavefunctions of the vector mesons. Three different types of vector-meson
wavefunction were studied for light mesons in \cite{fss:04a}. For the
$J/\Psi$ we shall use wavefunctions of exactly the same functional form, but
with the parameters adjusted to take into account the mass and charge of the
charmed quark and the experimental value of electronic decay width of the
$J/\Psi$-meson. In what follows, we shall comment briefly on the resulting
wavefunctions, referring to \cite{fss:04a} for detailed formulae and
discussion. 

In the DGKP approach \cite{dgkp:97}, the $r$ and $z$ dependence of the scalar
wavefunction is assumed to factorise into a product of gaussians. In the other
two cases considered in \cite{fss:04a}, it is obtained by taking a given
wavefunction in the meson rest frame. This is then boosted
into a light-cone wavefunction using the
Brodsky-Huang-Lepage prescription, in which the expressions for the
off-shellness in the centre-of-mass and light-cone frames are equated
\cite{bhl:80} (or equivalently, the expressions for the invariant mass of the
$q\bar{q}$ pair in the centre-of-mass and light-cone frames are equated
\cite{dgs:01}). In the simplest version of this approach, the wavefunction
assumes a gaussian form in the meson rest frame. Alternatively NNPZ
\cite{nnpz:97} have supplemented this by adding a hard \textquotedblleft
Coulomb\textquotedblright\ contribution in the hope of improving the
description of the rest-frame wavefunction at small $r$. However there are
theoretical problems with this latter wavefunction, as discussed in
\cite{fss:04a}, and we shall confine our discussion here to the DGKP and
\textquotedblleft Boosted Gaussian\textquotedblright\ wavefunctions.

The appropriate wavefunction parameters for the DGKP and Boosted Gaussian
wavefunctions in the $J/\Psi$ case are given, in the notation of
\cite{fss:04a}, in Tables \ref{tab:dgkp-parameters} and
\ref{tab:nnpz-parameters} respectively.\footnote{Compare with Tables 2, 3 and
4 of \cite{fss:04a}.} For a charm quark of given mass, they are chosen so that
the wavefunction is normalised and so that the predictions of the decay
constant are consistent with the experimental value $f_{J/\Psi}=0.273\pm0.005$ GeV
(see \cite{fss:04a} for further details).

\begin{table}[h]
\begin{center}
\textbf{DGKP parameters}
\[%
\begin{array}
[c]{|c|c|c|c|c|}\hline
m_{c} & \omega_{L} & \omega_{T} & \mathcal{N}_{L} & ~~~~\mathcal{N}%
_{T}\\\hline
1.4 & 0.688 & 0.560 & 18.941 & 8.280\\\hline
1.35 & 0.688 & 0.568 & 18.941 & 8.616 \\ \hline
\end{array}
\]
\end{center}
\caption{Parameters and normalisations of the DGKP light-cone wavefunctions in
appropriate GeV based units.}%
\label{tab:dgkp-parameters}%
\end{table}

\begin{table}[h]
\begin{center}
\textbf{Boosted Gaussian parameters}
\[%
\begin{array}
[c]{|c|c|c|c|c|c|}\hline
m_{c} & R^{2} & \mathcal{N}_{L} & \mathcal{N}_{T} & f_{V}(L) & f_{V}%
(T)\\\hline
1.4 & 2.44 & 0.0363 & 0.0365 & 0.262 & 0.293\\\hline
1.35 & 2.44 & 0.0369 & 0.0370 & 0.266 & 0.288 \\ \hline
\end{array}
\]
\end{center}
\caption{Parameters and normalisations of the Boosted Gaussian light-cone
wavefunctions in appropriate GeV based units.}
\label{tab:nnpz-parameters}%
\end{table}

The resulting wavefunctions are shown in Figs.\ref{fig:DGKPJpsi}--\ref{fig:gaussjpsi} 
for the case $m_{c}=1.4$ GeV. Like the corresponding
wavefunctions for the $\rho$ and $\phi$ mesons (shown in \cite{fss:04a}), 
the wavefunctions peak at
$z=0.5$ and $r=0$, and go to zero as $z\rightarrow0,1$ and $r\rightarrow
\infty$. As expected, for the $J/\Psi$ case the peaks are much sharper. We
also see that the DGKP and Boosted Gaussian wavefunctions are qualitatively
similar, with the transverse wavefunction having a broader distribution than
the longitudinal wavefunction in each case.

\subsection{$J/\Psi$ electroproduction}

\begin{figure}
\begin{center}
\includegraphics*[width=10cm]{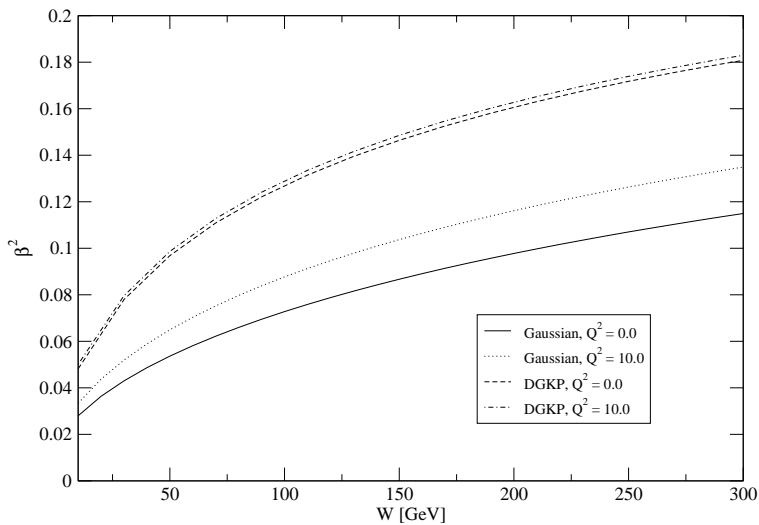}
\end{center}
\caption{The $W$ dependence of the squared ratio of the real to imaginary
parts of the $J/\Psi$ production amplitude (off transverse photons) obtained 
using the FS04 Regge dipole model.}
\label{Fig:betasqFSRegge_W}
\end{figure}

\begin{figure}
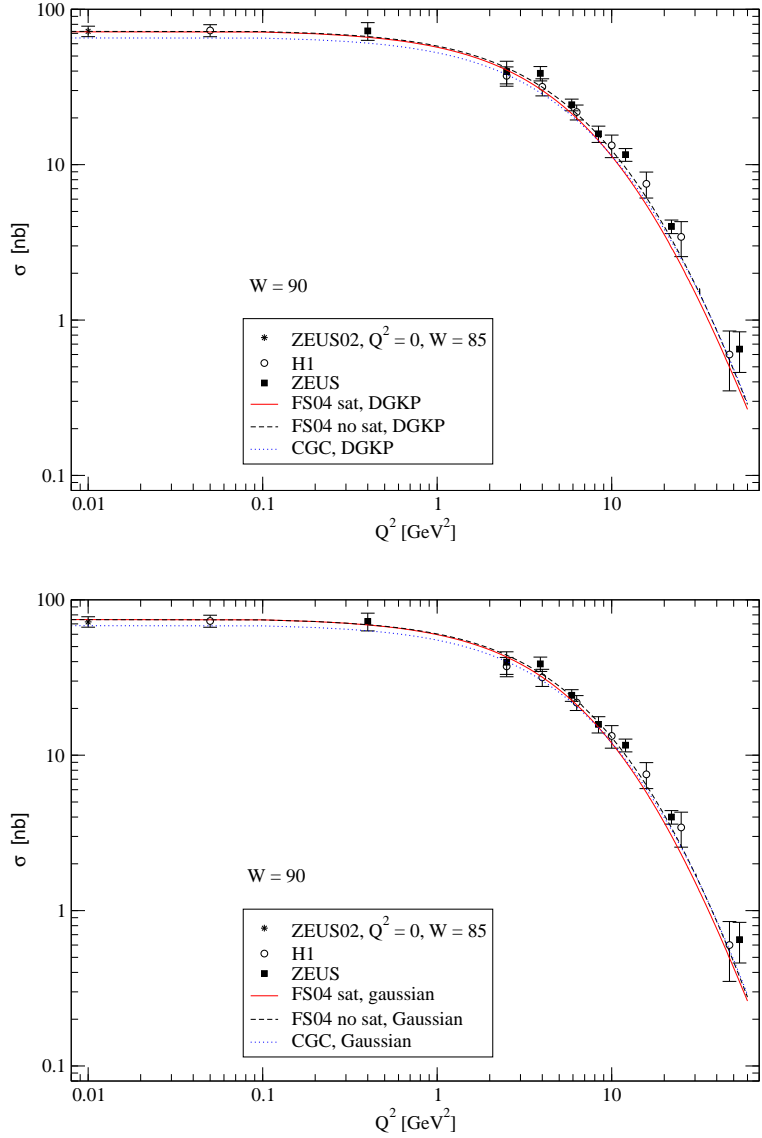

\begin{center}
\includegraphics*[width=10cm]{FIGS/jpsi_DGKP_q2.eps} \\ \vspace*{0.6cm}
\includegraphics*[width=10cm]{FIGS/jpsi_gauss_q2.eps}
\end{center}
\caption{Comparison of the model predictions to the data \cite{psi_H1,psi_ZEUS}
for exclusive $J/\Psi$ meson production: $Q^{2}$ dependence. 
Upper plot: DGKP wavefunction. Lower plot: Boosted Gaussian wavefunction.}
\label{fig:psiQ2}
\end{figure}

\begin{figure}[h]
\begin{center}
\includegraphics*[width=13cm]{FIGS/jpsi_DGKP_W1.eps} \\
\vspace*{1cm}
\includegraphics*[width=13cm]{FIGS/jpsi_DGKP_W2.eps}
\end{center}
\caption{Comparison of the model predictions to the data \cite{psi_H1,psi_ZEUS} 
for exclusive $J/\Psi$ meson production using the DGKP meson wavefunction: $W$ dependence.}
\label{fig:dgkppsiW}
\end{figure}

\begin{figure}[h]
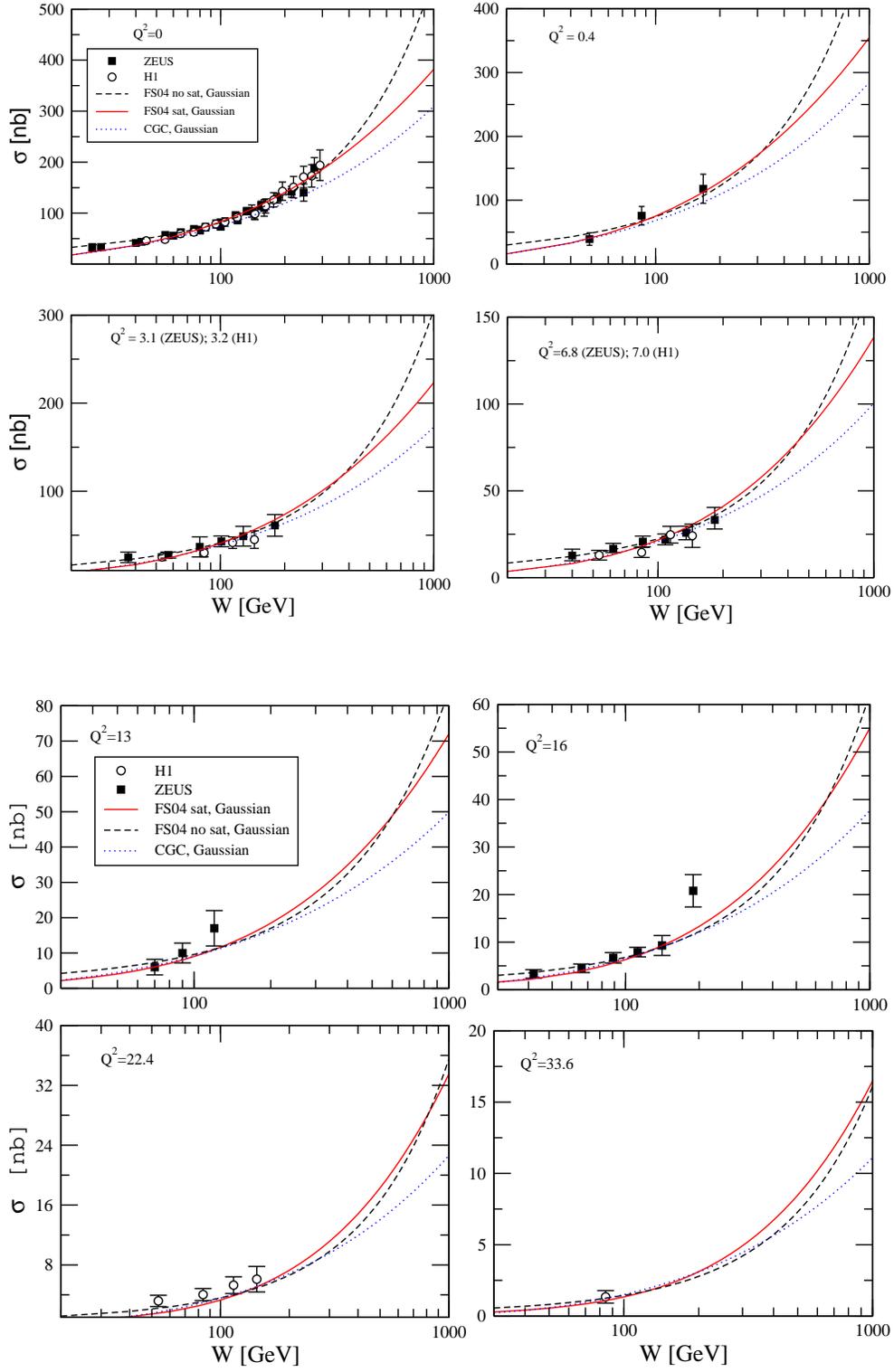

\begin{center}
\includegraphics*[width=13cm]{FIGS/jpsi_gauss_W1.eps} \\
\vspace*{1cm}
\includegraphics*[width=13cm]{FIGS/jpsi_gauss_W2.eps}
\end{center}
\caption{Comparison of the model predictions to the data \cite{psi_H1,psi_ZEUS}
for exclusive $J/\Psi$ meson production using the Boosted Gaussian meson wavefunction: $W$
dependence.}
\label{fig:gausspsiW}
\end{figure}

\begin{figure}
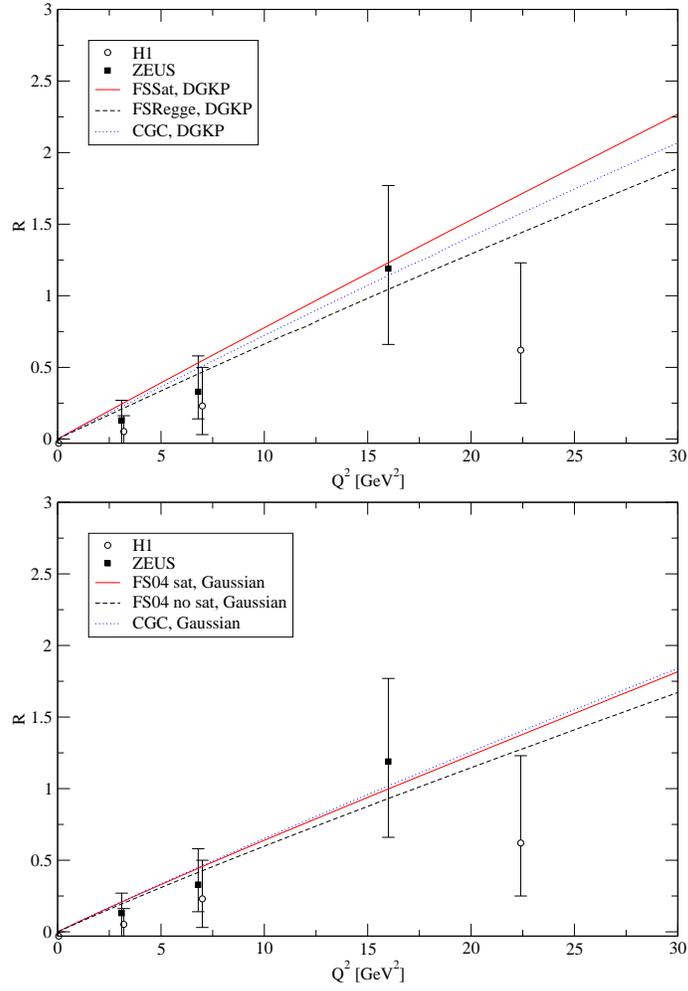

\begin{center}
\includegraphics*[width=9cm]{FIGS/jpsi_DGKP_ratio_q2.eps} \\
\includegraphics*[width=9cm]{FIGS/jpsi_gauss_ratio_q2.eps}
\end{center}
\caption{Comparison of the model predictions to the data \cite{psi_H1,psi_ZEUS}
on the ratio $R$ for exclusive $J/\Psi$ meson.}
\label{fig:Rpsi}
\end{figure}

Given the forms of the photon and $J/\Psi$ wavefunctions, the cross-section
in Eq.(\ref{sigmatot}) can be calculated from 
Eqs.(\ref{result1})--(\ref{sigmatot}).
However, to do this requires an estimate of the correction due to the real
part in Eq.(\ref{fdcs}) and of the slope parameter $B$ in Eq.(\ref{sigmatot}).

The correction factor $(1+\beta^2)$ for the real part of the amplitude is, 
as for the DVCS case, determined from the
FS04 Regge model. The resulting values are 
illustrated in Fig.\ref{Fig:betasqFSRegge_W}, which shows
the $\beta^{2}$ values obtained as functions of $W$ at $Q^{2}=0,\,10$
GeV$^{2}$. As can be seen, the corrections from the real parts in 
Eq.(\ref{fdcs}) are a significant, but not large,
correction. For the slope parameter $B$ in Eq.(\ref{sigmatot}) 
we use the simple parameterisation (in GeV units)
\begin{equation}
B=N\left(  \frac{14.0}{\left(  Q^{2}+M_{V}^{2}\right)  ^{0.3}}+1\right)
\label{Bslope}
\end{equation}
with $N=0.55$ GeV$^{-2}$ for the $J/\Psi$ which is in accord with
the data \cite{psi_H1,psi_ZEUS}.

Having fixed the real parts and the slope parameter, predictions for the
production cross-section can be made for both the Boosted Gaussian and DGKP
wavefunctions. The charmed quark mass is adjusted within the range allowed by
the charm structure function data, i.e. we use $m_{c}=1.4\,$\ GeV for the
FS04 Regge model and $m_{c}=1.35$ GeV for the two saturation models. 

The predictions for the $Q^2$ dependence of the total cross-section 
$\sigma^{\text{TOT}}=\sigma^{\text{T}} +\epsilon\sigma^{\text{L}}$ (we take $\epsilon=0.98$) 
are presented in Fig.\ref{fig:psiQ2}. There is very good agreement 
between theory and data with relatively little dependence upon the choice of meson 
wavefunction.\footnote{Unlike the case for the light mesons \cite{fss:04a}.}
We note that for $Q^2 \lesssim 5$ GeV$^2$, the predictions are very sensitive to the 
choice of charm quark mass. For example, the small difference between the
predictions of the CGC and the other two model predictions in this region can
be eliminated by fine-tuning the charm quark mass in the CGC case from 1.35 to
1.32 GeV. 

The $W$-dependence is shown for a range of $Q^{2}$ values in 
Figs.\ref{fig:dgkppsiW}--\ref{fig:gausspsiW}. As before, agreement is good
and there is little dependence on the choice 
of meson wavefunction. Large differences between the models only arise at 
energies beyond the current experimental range ($W\gtrsim400$ GeV), but again
the differences between the two saturation models can be significantly reduced
by fine-tuning 
the chosen values for the charmed quark mass, especially at low $Q^2$.  

Finally, our predictions for the $Q^{2}$ dependence of the cross-section ratio
$R=\sigma_{L}/\sigma_{T}$ at $W=90$ GeV are shown in Fig.\ref{fig:Rpsi}. There is once
again good agreement between theory and data with slightly more dependence upon
the choice of meson wavefunction. Note that the longitudinal-to-transverse
ratio $R$ increases approximately linearly with $Q^{2}$ and, in contrast to
the $\rho$ and $\phi$ cases \cite{fss:04a}, does not flatten out in the currently accessible
range of $Q^{2}$. This is to be expected, since in the extreme
non-relativistic limit the wavefunction approaches a delta function whence the
ratio $R\propto Q^{2}/m_{c}^{2}$ for all $Q^{2}$.

\clearpage

\section{Diffractive deep inelastic scattering (DDIS)}

\begin{figure}[h]
\begin{center}
\includegraphics[width=5cm]{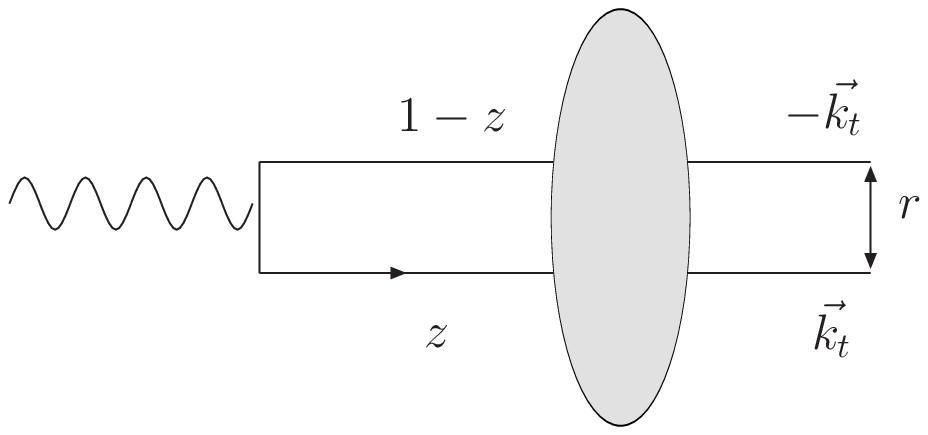}
\includegraphics[width=5cm]{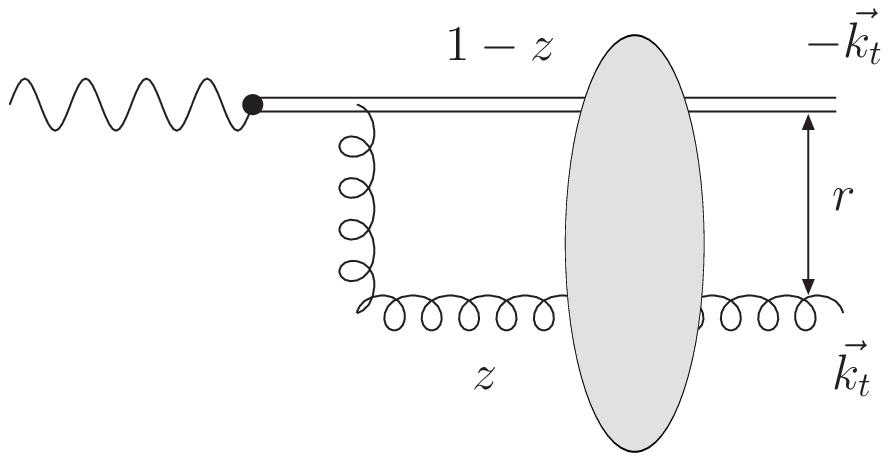}
\end{center}
\caption{The $q\bar{q}$ and $q\bar{q}g$ contributions to $F_{2}^{D(3)}$.}
\label{fig:feynman}
\end{figure}

To conclude our study we turn to the diffractive deep inelastic scattering
(DDIS) process
$$ \gamma^{\ast}+p\rightarrow X+p\;\;,\label{DDIS} $$
where the hadronic state $X$ is separated from the proton by a rapidity gap.
In this process, in addition to the usual variables $x$ and $Q^{2}$ there is a
third variable $M_{X}^{2}$. In practice,
$x$ and $M_{X}^{2}$ are often replaced by the variables $x_{I\!\!P}$ and
$\beta$:
\begin{equation}
x_{I\!\!P}\simeq\frac{M_{X}^{2}+Q^{2}}{W^{2}+Q^{2}}\hspace{1cm}\beta=\frac
{x}{x_{I\!\!P}}\simeq\frac{Q^{2}}{M_{X}^{2}+Q^{2}}~.
\label{eq:diffractive.variables}
\end{equation}
In the diffractive limit $s\gg Q^{2},m_{X}^{2}$ and so $x_{I\!\!P}\ll1$.

\begin{figure}
\begin{center}
\includegraphics*[width=15cm]{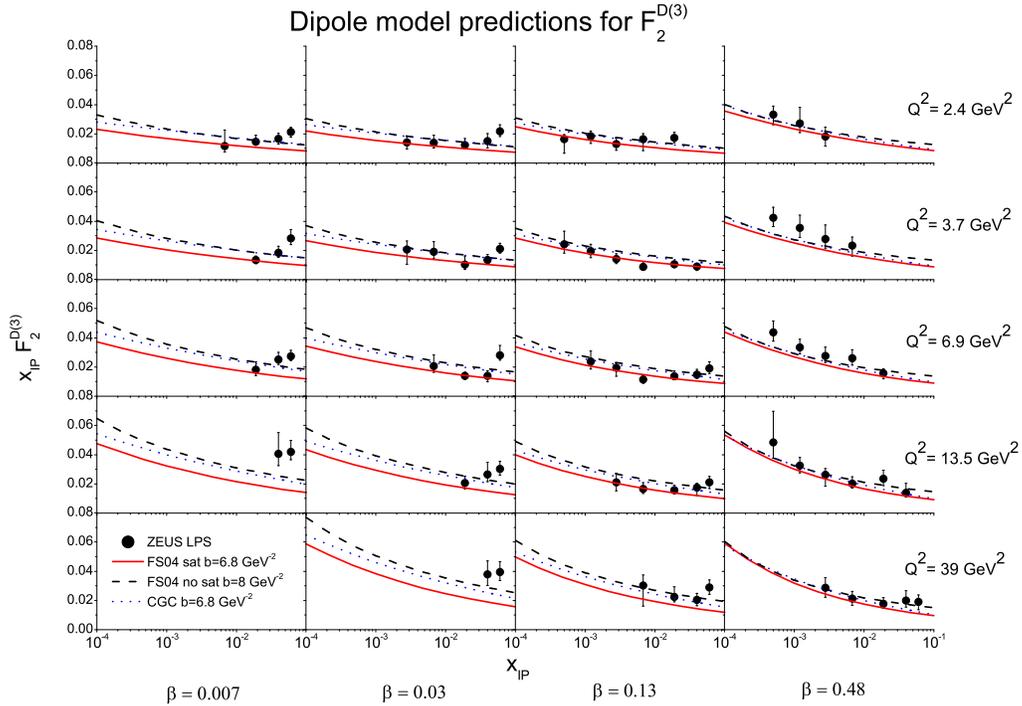}
\end{center}
\caption{Model predictions compared to the ZEUS LPS data \cite{LPS05}.}
\label{fig:F2D3LPS}
\end{figure}

\begin{figure}
\begin{center}
\includegraphics*[width=16cm]{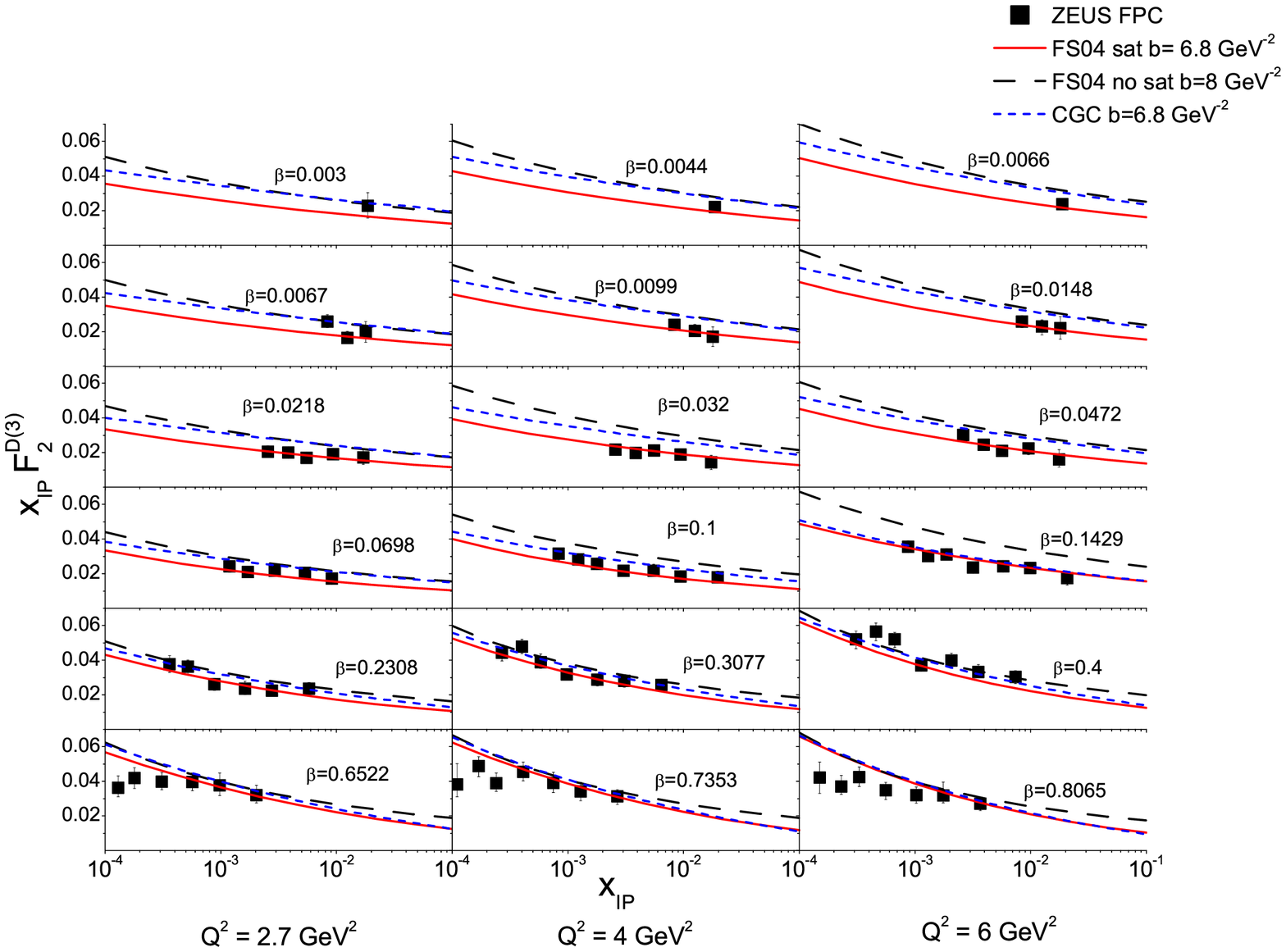}
\end{center}
\caption{Model predictions compared to the ZEUS FPC data 
with $M_Y < 2.3$ GeV (low $Q^{2}$) \cite{FPC05}.}
\label{fig:F2D3FPC1}
\end{figure}\begin{figure}
\begin{center}
\includegraphics*[width=16cm]{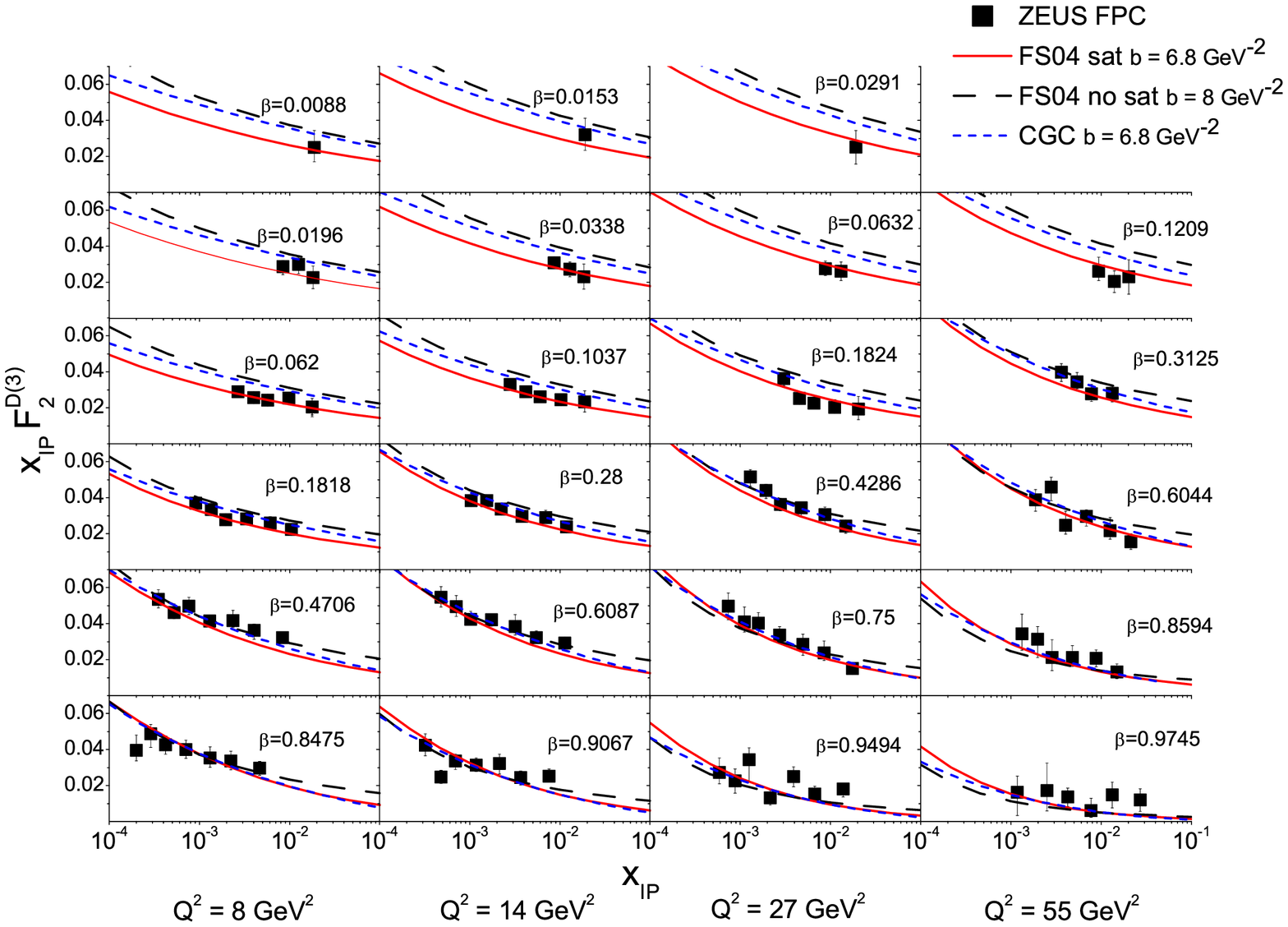}
\end{center}
\caption{Model predictions compared to the ZEUS FPC data with
$M_Y < 2.3$ GeV (high $Q^{2}$) \cite{FPC05}.}
\label{fig:F2D3FPC2}
\end{figure}

\begin{figure}
\begin{center}
\includegraphics*[width=15cm]{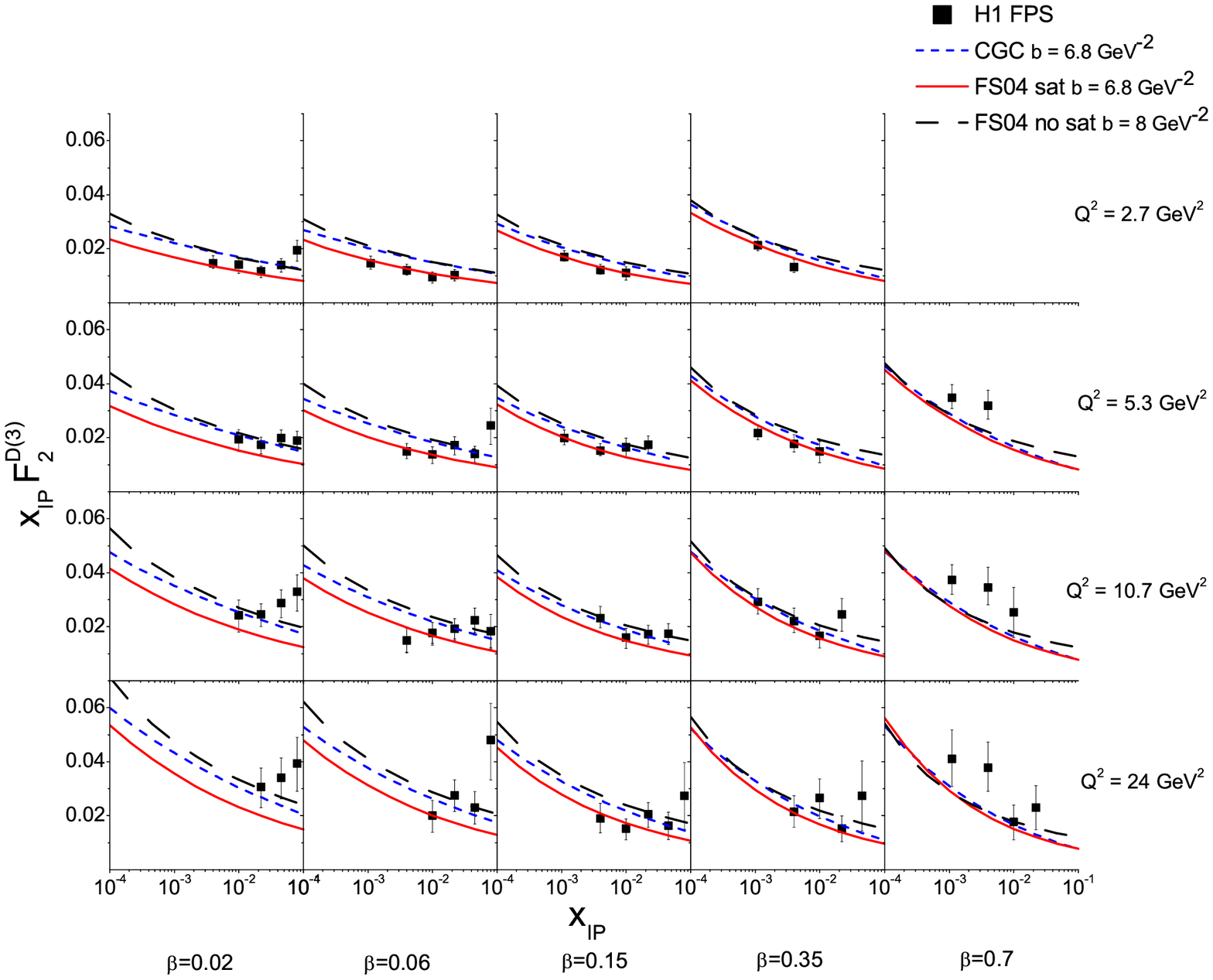}
\end{center}
\caption{Model predictions compared to the H1 FPS data: $\xp$ dependence \cite{H1FPS}.}
\label{fig:F2D3FPS1}
\end{figure}
\begin{figure}
\begin{center}\includegraphics*[width=13.5cm]{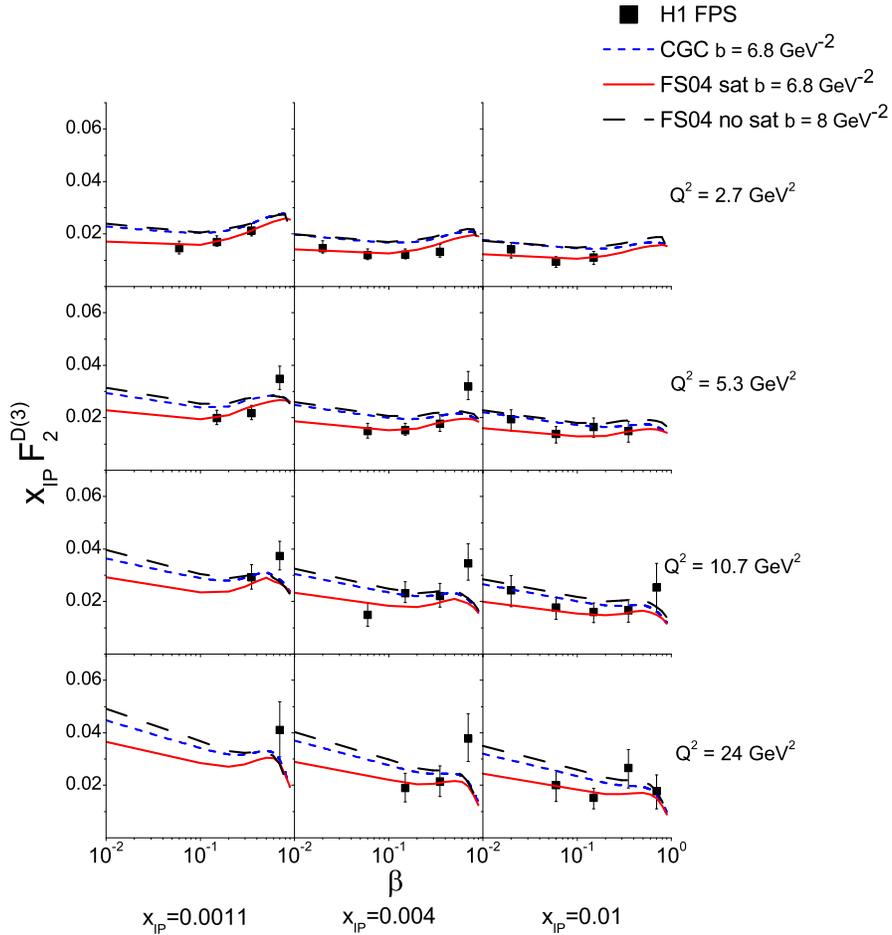}
\end{center}
\caption{Model predictions compared to the H1 FPS data: $\beta$ dependence \cite{H1FPS}.}
\label{fig:F2D3FPS2}
\end{figure}

\begin{figure}
\begin{center}
\includegraphics*[width=13.5cm]{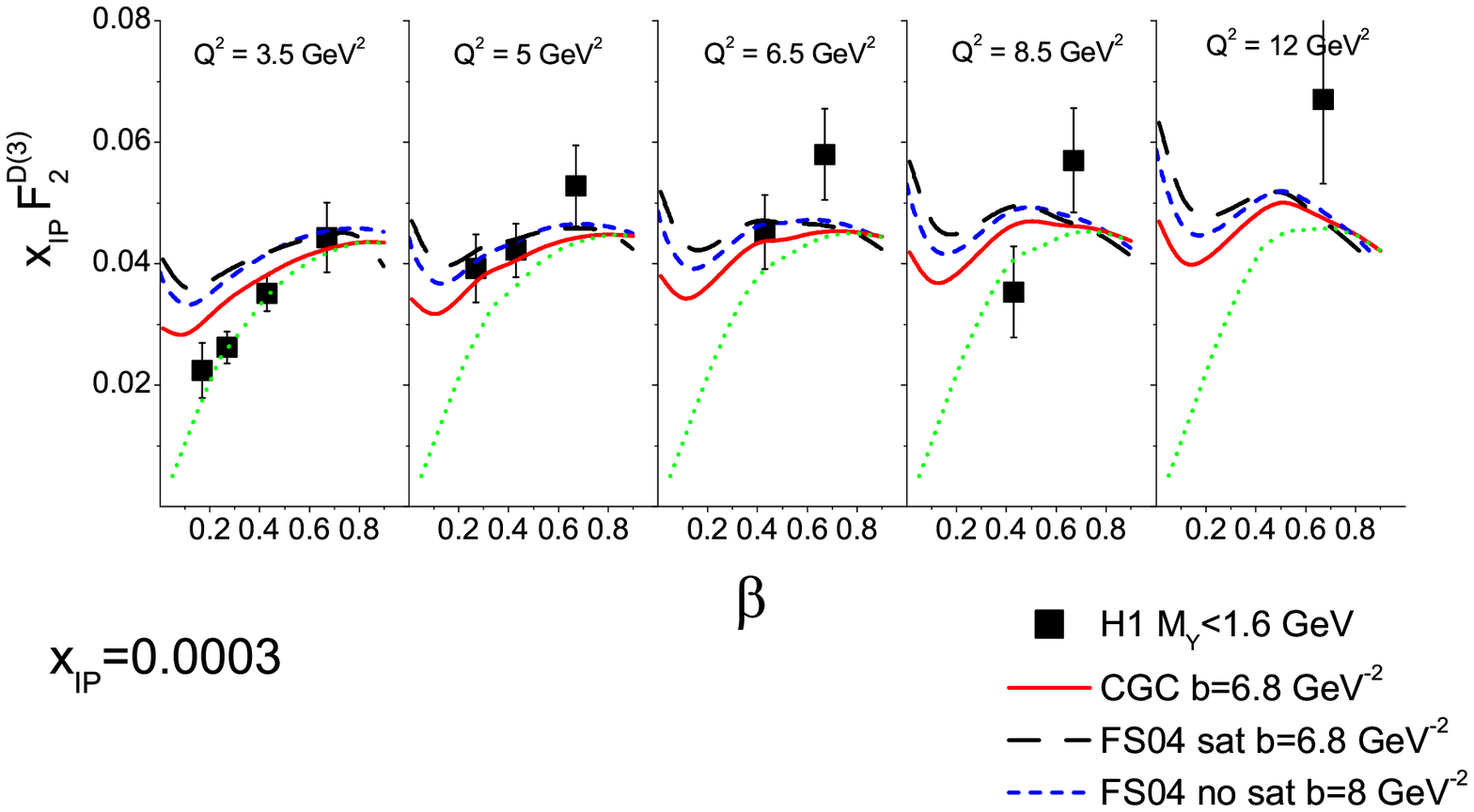}
\end{center}
\caption{Model predictions compared to the H1 data with $M_{Y}<1.6$ GeV \cite{H1MY}: 
$\beta$ dependence at $\xp=0.0003$. 
Green dotted curve shows the contribution without including the $q\bar{q}g$ component.
}
\label{fig:F2D3H1beta1}
\end{figure}

\begin{figure}
\begin{center}
\includegraphics*[width=13.5cm]{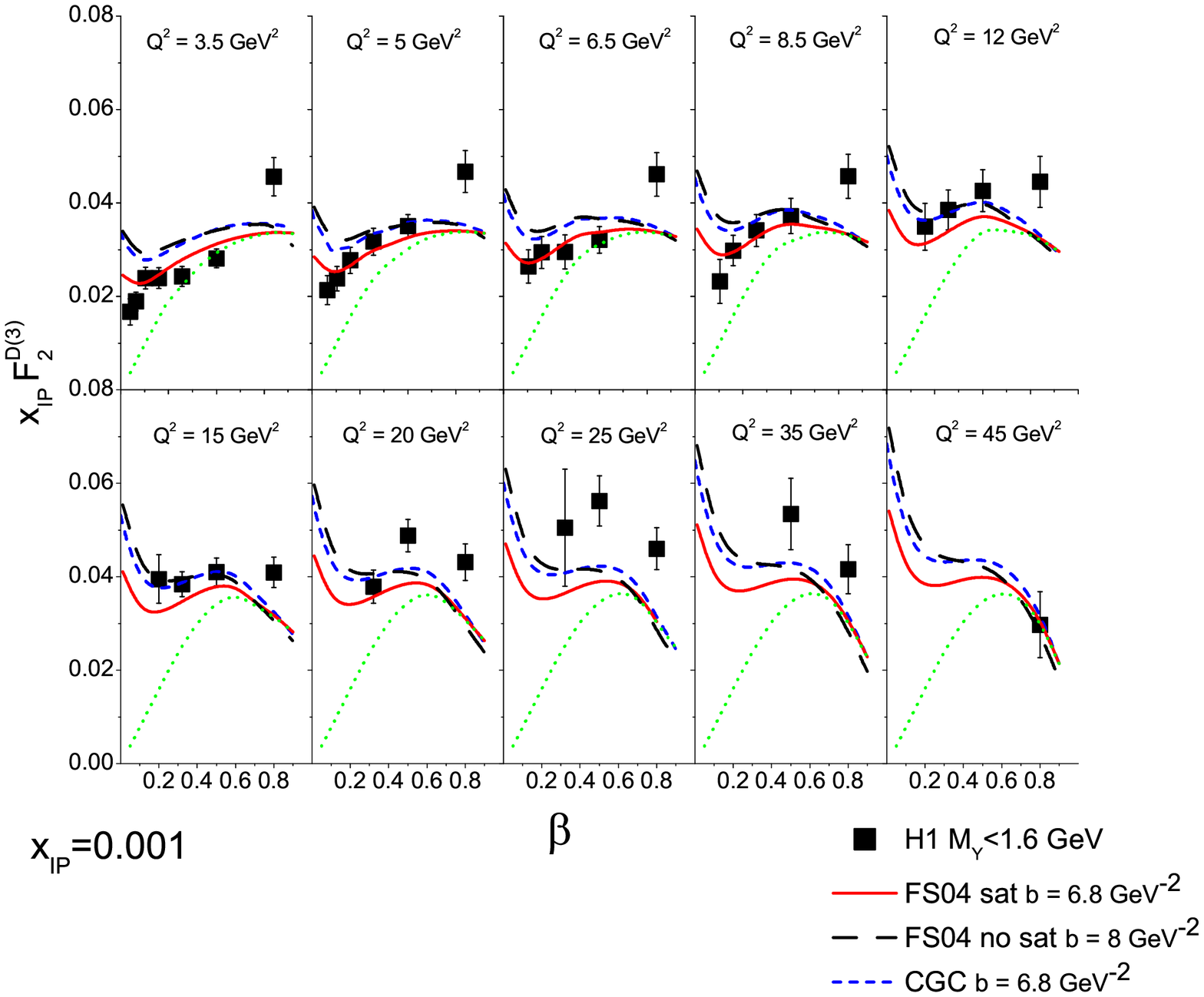}
\end{center}
\caption{Model predictions compared to the H1 data with $M_{Y}<1.6$ GeV \cite{H1MY}: 
$\beta$ dependence at $\xp=0.001.$
Green dotted curve shows the contribution without including the $q\bar{q}g$ component.
}
\label{fig:F2D3H1beta2}
\end{figure}

\begin{figure}
\begin{center}
\includegraphics*[width=13.5cm]{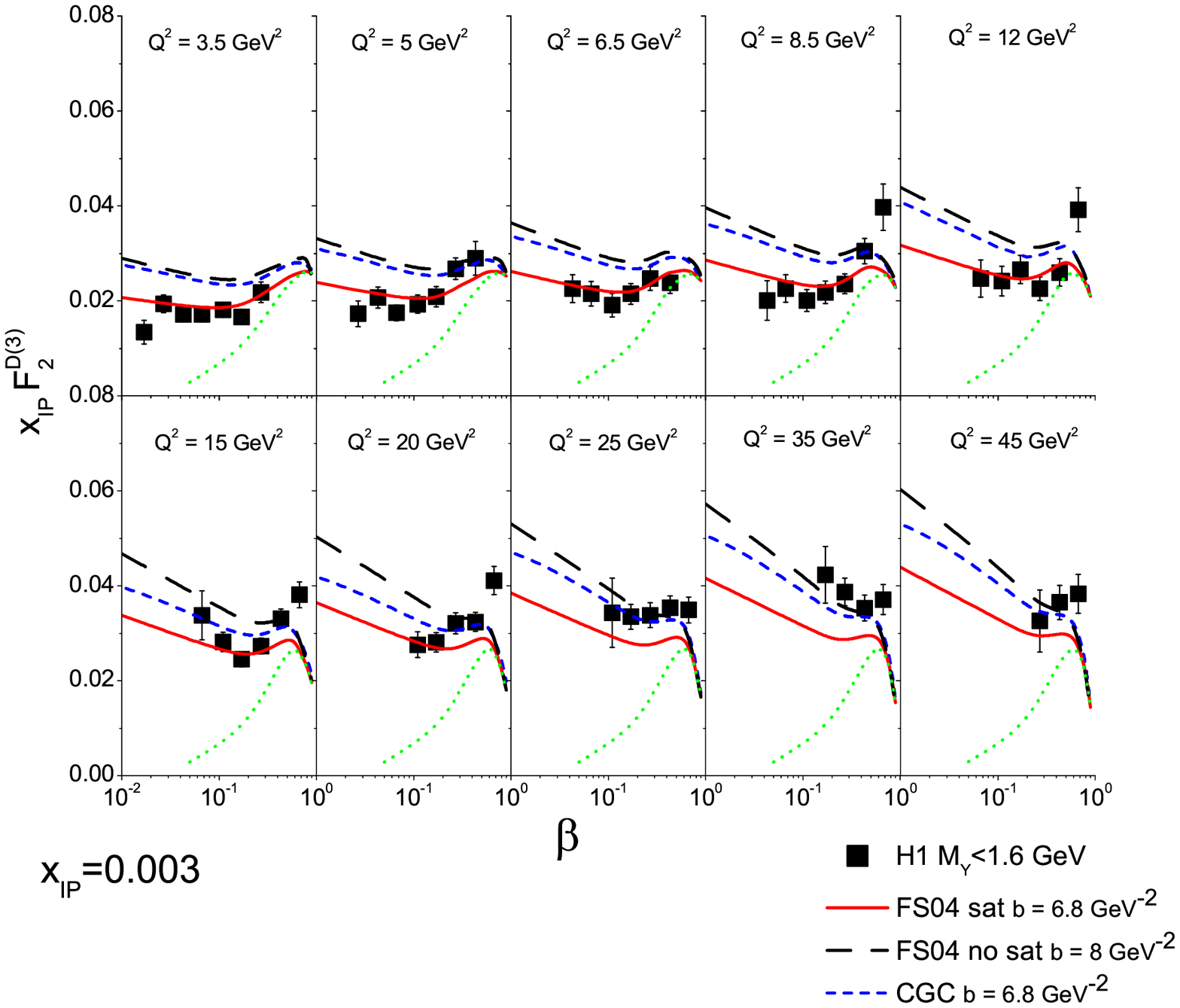}
\end{center}
\caption{Model predictions compared to the H1 data with $M_{Y}<1.6$ GeV \cite{H1MY}: 
$\beta$ dependence at $\xp=0.003.$
Green dotted curve shows the contribution without including the $q\bar{q}g$ component.
}
\label{fig:F2D3H1beta3}
\end{figure}

\begin{sidewaysfigure}
\begin{center}
\includegraphics[width=20cm]{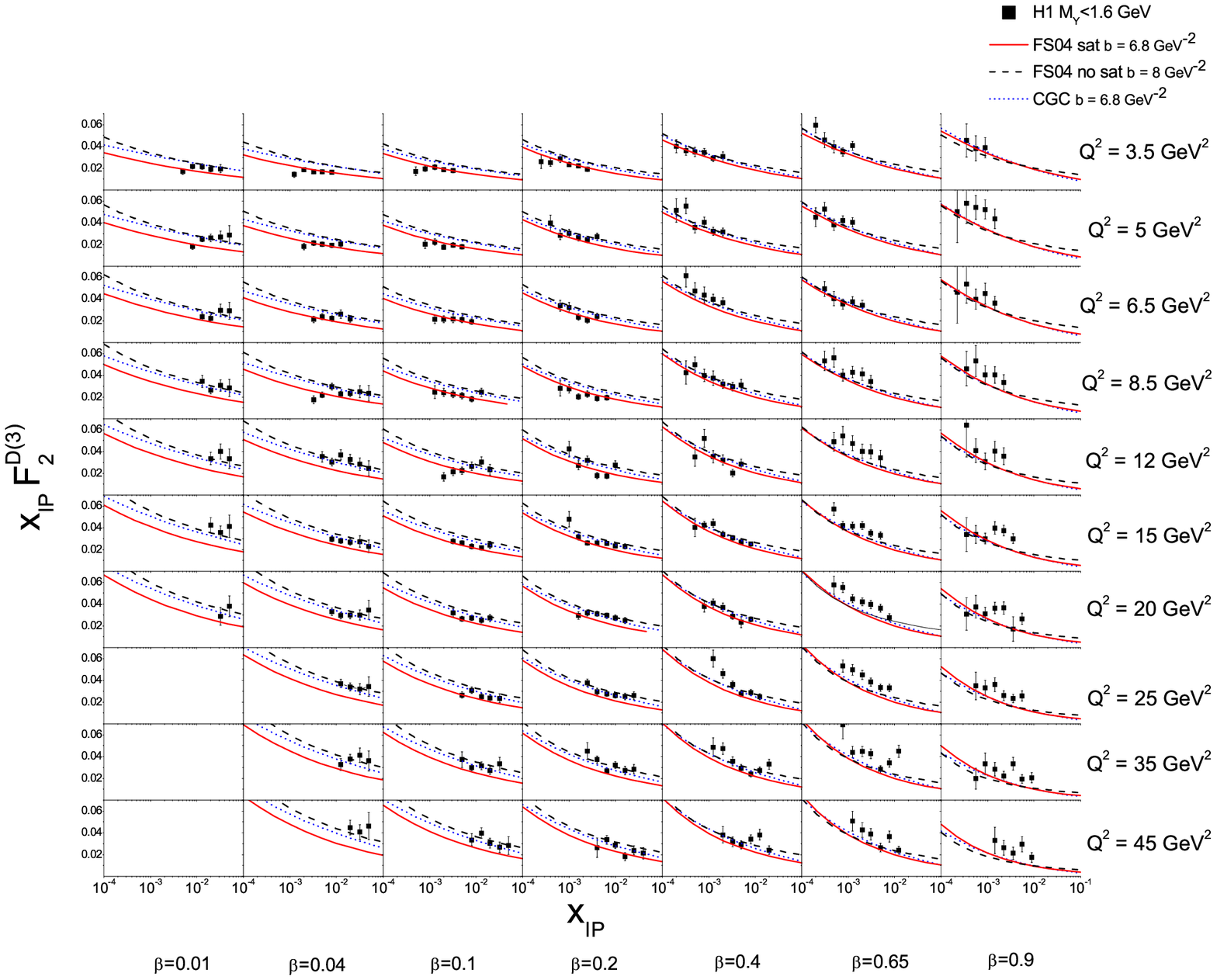}
\end{center}
\caption{Model predictions compared to the H1 data with $M_{Y}<1.6$ GeV \cite{H1MY}: 
$x_{I\!\!P}$ dependence.}
\label{fig:F2D3H1xpom}
\end{sidewaysfigure}

In the dipole model, the contribution due to quark-antiquark dipoles to the
structure function $F_{2}^{D(3)}$ can be obtained from a momentum space
treatment as described in \cite{Wusthoff97,GW99a}. However, if we are to
confront the data at low values of $\beta$, corresponding to large invariant
masses $M_{X}$, it is necessary also to include a contribution from the higher
Fock state $q\bar{q}g$. We can estimate this contribution using an effective
\textquotedblleft two-gluon dipole\textquotedblright\ approximation due to
W\"{u}sthoff \cite{Wusthoff97}, as illustrated in Fig.\ref{fig:feynman}.

Again, the predictions obtained in this way involve no adjustment of the
dipole cross-sections and photon wavefunctions used to describe the $F_{2}$
data. We are however free to adjust the forward slope for inclusive
diffraction ($B$) within the range acceptable to experiment, which means that
the overall normalisation, but not the energy dependence, of $F_{2}^{D(3)}$ is
free to vary somewhat. We take $B=6.8$ GeV$^{-2}$ when making our 
CGC and FS04 saturation predictions and $B=8.0$ GeV$^{-2}$ when 
making the FS04 Regge predictions. Note that a value of $8.0$ GeV$^{-2}$ is rather
high compared to the $\approx 6$ GeV$^{-2}$ favoured by the H1 FPS data \cite{H1FPS}
although it is in the range allowed by the ZEUS LPS data \cite{LPS05}. 
The need for a larger value of $B$ for the FS04 Regge model arises since the
corresponding dipole cross-section is significantly larger than the FS04 saturation
model at large values of $r$ and this enhancment is magnified in inclusive 
diffraction since it is sensitive to the square of the dipole cross-section.
We should also bear in mind that the tagged proton data are subject to an overall 
$\approx 10\%$ normalisation uncertainty. We are also somewhat free to vary 
the value of $\alpha_{s}$ used to define the normalisation of the model 
dependent $q\bar{q}g$ component, which is important at low values of $\beta$. Rather
arbitrarily we take $\alpha_{s}=0.1$ and take the view that the theory curves
are less certain in the low $\beta$ region.

In Fig.\ref{fig:F2D3LPS} we compare the recent ZEUS LPS data \cite{LPS05} on
the $x_{I\!\!P}$ dependence of the structure function $F_{2}^{D(3)}$ at
various fixed $Q^{2}$ and $\beta$ with the models\footnote{Predictions for the
original Iancu et al CGC model have previously been published in
\cite{fss:04b}.}. The agreement is good except at the larger $x_{I\!\!P}$
values. Indeed, the $\chi^2$ values per
data point are very close to unity for all three models for $\xp < 0.01$. 
Disagreement at larger $\xp$ is to be expected since this is the 
region where we anticipate a significant non-diffractive contribution which is absent 
in the dipole model prediction. Note that the three models produce similar 
predictions at larger values of $\beta$.

Contamination from secondary exchanges is avoided in the FPC data
\cite{FPC05}, in which the non-diffractive contribution is explicitly removed
by the ZEUS method of analysis. However this comes at the cost of including
proton dissociation contributions, since the mass of the target fragments ($M_Y$) 
is only limited to being below 2.3 GeV. The predictions of our three models are
compared with these data in Fig.\ref{fig:F2D3FPC1} and Fig.\ref{fig:F2D3FPC2},
where the theory curves have been divided by $0.7$ to allow for proton
dissociation contributions, assumed to be a fixed fraction of the total,
independent of the kinematic variables. This is obviously a crude
approximation but it is supported by the experimental analyses.
These data suggest that the CGC and Regge model 
predictions are perhaps overshooting the data at low $\beta$ whilst the three 
models produce very similar predictions at larger values of $\beta$, 
i.e. $\beta \gtrsim 0.4$. Specifically, for the FS04 saturation model $\chi^2=238$ for
the entire 156 data points whilst the CGC and FS04 Regge models have $\chi^2 = 147$ and 
150 respectively for the 94 data points with $\beta > 0.3$.
However, we must be careful not to overinterpret the data.
The low $\beta$ region is precisely the region where there is an appreciable
$q\bar{q}g$ component which is subject to a rather large theoretical
uncertainty. Moreover, we must not forget that (for the FPC data) we have 
assumed that the fraction of events which contain a dissociated proton is 
constant. Recall also that there is an uncertainty in the value of the forward slope
parameter used in determining the theoretical predictions.

Comparison to the H1 data with tagged protons \cite{H1FPS} is to be found in 
Fig.\ref{fig:F2D3FPS1} and Fig.\ref{fig:F2D3FPS2}. The story is similar to that for
the ZEUS data and the evidence for an overshoot of the CGC and Regge
model predictions at low $\beta$ is strengthened. The $Q^2=2.7$ GeV$^2$ panes 
in Fig.\ref{fig:F2D3FPS2} illustrate this point the best. Again, we should
not interpret this as evidence against these dipole models due to the uncertainty in
the $q\bar{q}g$ contribution in the low $\beta$ region. The agreement between all
models and the data at larger values of $\beta$ and low enough $\xp$
is satisfactory.\footnote{The FS04 saturation fit has $\chi^2=37$ for the 40 points 
with $\xp < 0.01$ whilst the CGC and FS04 Regge fits both have $\chi^2$ per data point 
below 2 for the 18 points with $\beta \ge 0.35$.}

In Figs.\ref{fig:F2D3H1beta1}--\ref{fig:F2D3H1xpom} we compare to the H1 data collected 
with some proton dissociation, i.e. $M_{Y}<1.6$ GeV \cite{H1MY}, and in order to include 
the effect of proton dissociation we now divide the theory by a factor $0.8$.
The FS04 saturation model performs well over the whole kinematic region with
$\xp \lesssim 10^{-2}$ whilst once again the CGC and Regge models overshoot the data 
at low $\beta$. On these plots we also show the result of computing $F_2^{D(3)}$ without
including the $q\bar{q}g$ component: clearly the data require an additional contribution at
low $\beta$. We note that it is possible to improve the quality of the agreement between
data and theory if we are allowed to increase the fraction of proton dissociation assumed
in the data (there is perhaps a hint that this increase should also be slightly larger at 
higher values of $Q^2$).
For example, for the 170 data points in Fig.\ref{fig:F2D3H1xpom} with $\beta \ge 0.4$
and $\xp < 0.01$ we find $\chi^2$ values of 244, 330 and 263 for the FS04 saturation,
FS04 Regge and CGC models respectively if we globally decrease the scaling factor from 
0.8 to 0.7. 

Before leaving the H1 data we should say that, strictly speaking, all of the 
H1 data refer to the reduced diffractive cross-section, $\sigma_r^{D(3)}$. 
However the difference between that
quantity and the diffractive structure function $F_2^{D(3)}$ is mostly 
negligible and is never more than $10\%$.  

In summary, the DDIS data at large enough $\beta \gtrsim 0.4$ and small enough 
$\xp \lesssim 0.01$ are consistent with the predictions of all three dipole 
models. However the data themselves would have a much greater power to
discriminate between models if the forward slope parameter were measured to
better accuracy. At smaller values of $\beta$, the data clearly reveal the 
presence of higher mass diffractive states which can be estimated via the
inclusion of a $q\bar{q}g$ component in the dipole model calculation
under the assumption that the three-parton system interacts as a single dipole
according to the universal dipole cross-section. The theoretical calculation
at low $\beta$ must be improved before the data in the region can be utilised 
to disentangle the physics of the dipole cross-section. Nevertheless, it is
re-assuring to observe the broad agreement between theory and data in the low
$\beta$ region.    

\section{Conclusion}
The dipole scattering approach, when applied to diffractive 
electroproduction processes, clearly works very well indeed. The HERA
data now constitute a large body of data which is typically accurate to 
the 10\% level or better, and without exception the dipole model is able
to explain the data in terms of a single universal dipole scattering
cross-section. Perhaps the most important question to ask of the data is
the extent to which saturation dynamics is present. Although the $F_2$ 
data suggest the presence of saturation dynamics \cite{FS04}, 
the remaining data on exclusive processes and on $F_2^{D(3)}$ 
are unable to distinguish between the models we consider here:
these data are therefore unable to offer additional information on the possible
role of saturation. We do note that a more accurate determination of the forward
slope parameter in diffractive photo/electro-production processes would 
significantly enhance the impact of the data. However, it is hard to avoid the
conclusion that only with more precise data or with data out to larger values of 
the centre-of-mass energy will we have the chance to make a definitive statement on 
the role of saturation without the inclusion of the low $Q^2$ $F_2(x,Q^2)$ data in
the analysis.

\section*{Acknowledgments}
This research was supported in part by the UK's Particle Physics and Astronomy
Research Council. We should like to thank Paul Laycock and Paul Newman for
helpful discussions.

\end{document}